\documentclass[final,3p,times,authoryear]{elsarticle}
\usepackage[utf8]{inputenc}

\usepackage{graphicx}
\usepackage{tikz}

\usepackage{pgfplots}
\pgfplotsset{compat=newest}
\usetikzlibrary[pgfplots.groupplots]
\usetikzlibrary{arrows}
\usetikzlibrary{decorations.markings}

\usetikzlibrary{backgrounds}
\usetikzlibrary{chains}
\usetikzlibrary{calc}
\usetikzlibrary{external}
\usepackage{bm}
\usepackage{hyperref}

\usepackage{empheq}
\usepackage{xspace}
\usepackage{color}
\usepackage{amssymb}
\usepackage{booktabs}

\hypersetup{
bookmarksnumbered = true,
unicode=false,          
pdftoolbar=true,        
pdfmenubar=true,        
pdffitwindow=false,     
pdfstartview={Fit},    
pdftitle={Bi-stability resistant to fluctuations},    
pdfauthor={M. Caruel},     
pdfsubject={Subject},   
pdfcreator={Creator},   
pdfproducer={Producer}, 
pdfkeywords={keyword1} {key2} {key3}, 
pdfnewwindow=true,      
colorlinks=true,       
linkcolor=black,          
citecolor=black,        
filecolor=black,      
urlcolor=black,           
pdfdisplaydoctitle = true
}

\renewcommand{\bar}{\overline}

\renewcommand{\bar}[1]{\overline{#1}}

\newcommand{\of}[1]{\left(#1\right)}
\newcommand{\mean}[1]{\langle #1 \rangle}
\newcommand{\sech}{\mathrm{sech}}

\usepackage[normalem]{ulem}
\usepackage{color}
\definecolor{myorange}{rgb}{0.9568,0.4941,0.1961}
\definecolor{myred}{rgb}{0.9098,0.1294,0.2078}
\definecolor{myblue}{rgb}{0.0352,0.4981,0.6509}
\definecolor{mygreen}{rgb}{0.2235,0.6353,0.2588}

\usepackage{marvosym}

\journal{Journal of the Mechanics and Physics of Solids}
\begin{document}

\title{Bi-stability resistant to fluctuations}

\author[upec]{M. Caruel\corref{cor}}
\address[upec]{
Universit\'e Paris Est, Laboratoire Mod\'elisation et Simulation Multi Echelle, MSME-UMR-CNRS-8208, 61 Avenue du G\'en\'eral de Gaulle, 94010 Cr\'eteil, France
}
\ead{matthieu.caruel@u-pec.fr}
\cortext[cor]{Corresponding author}
\author[espci]{ L. Truskinovsky}
\address[espci]{
PMMH, CNRS UMR 7636, ESPCI, PSL, 10 rue de Vauquelin, 75231 Paris cedex 05, France
}
\date{\today}

\begin{abstract}
We study a simple micro-mechanical device that does not lose  its snap-through behavior in an environment dominated by  fluctuations. The main idea is to have several  degrees of freedom that can cooperatively  resist the  de-synchronizing effect of random perturbations.  As an inspiration  we use the  power stroke machinery of skeletal muscles, which ensures at sub-micron scales and  finite temperatures  a swift  recovery of  an abruptly applied slack.   In addition to  hypersensitive response at finite temperatures, our  prototypical  Brownian snap spring also exhibits criticality at special values of parameters which is another potentially  interesting property for micro-scale engineering applications. 
\end{abstract}

\maketitle

\section{Introduction} 
\label{sec:introduction}

Recent progress in micro and nano fabrication techniques opened new ways of  building materials endowed with non-conventional mechanical properties \citep{Nicolaou:2012cf,Florijn:2014fw,Haghpanah:2016ih,Cho:2016kh,Xin:2016by}.
  A large class of such  material can be viewed  as multi-scale  distributed mechanisms capable of 
internal switching transitions due to  controllable  micro instabilities.  The applications of the materials-mechanisms with tunable multi-stability  range from  energy harvesting and  acoustic cloaking to mechanical memory and tissue engineering \citep{Harne:2013if,Halg:1990fq,Receveur:2005eu,Nadkarni:2016jn,Harne:2016gg,Paulose:2015hda,Rafsanjani:2015fh,Buckmann:2015cr,Frenzel:2016hg}. 

Typical systems of this type require for their functioning mechanical  elements  with  nonconvex  elastic energy \citep{Shan:2015fz}, but it is not obvious that  such nonconvexity can be sustained at finite temperatures.  In other words, it is not clear  whether  the  discontinuous response of these  elements can survive progressive  miniaturization  to scales where thermal [or non-thermal \cite{Sheshka:2016dm}]  fluctuations cannot be neglected \citep{Steuerman:2004fx,Flood:2004kr,Ollier:1995hi,Li:2016hs}.  This question is relevant because at sufficiently high temperatures mechanical barriers can be crossed uncontrollably causing the  effective ``disintegration'' 
of the underlying snap-through mechanisms.
 Given that  suppression of  fluctuations at small scales may be either impossible, or too costly, special engineering solutions are needed that  take advantage of the fluctuations without  compromising the  robustness of the switching.

In search of how to build a mechanical snap spring that  preserves its hypersensitivity to loading in the presence of random fluctuations, we  turn to  biological systems which are  known to  operate with remarkable robustness in the  Brownian environment. For instance, muscle contraction, hair cell gating, cell adhesion and neurotransmitter release, all involve fast conformational transitions  which ensure their extremely sharp, almost digital (all-or-none) response to external mechanical  stimulation \citep{Sudhof:2013fw,Howard:1988uu,Martin_2000,Reconditi_2004}. Even a superficial examination shows that all these systems employ the same basic design principle to fight  the  de-synchronizing effect of temperature.

Since a system with one degree of freedom cannot resist thermal rounding of its mechanical response  \citep{Shiino:1987tc},
 a way to maintain multi-stability at elevated temperatures is to engage  simultaneously  many coupled degrees of freedom that can synchronize.
 To maintain the complexity of the energy landscape  in a fluctuating environment one  needs to ensure that the  energy barriers  separating different wells diverge with the system size.   This can happen when a system develops macro-wells corresponding to synchronized configurations involving large number of elements and when individual elements  can escape from such energy wells only  cooperatively. 
 
Biological examples suggest that this fundamental mechanism is ubiquitous in living systems. In inanimate world, similar idea is  behind the functioning  of ferromagnetic materials, where individual bi-stable elements (spins)  behave cooperatively at low temperatures, allowing one to store non-mechanical information robustly.
In such spatially extended systems, the switch  takes place in the form of  propagating  domain walls. 
A zero dimensional macro-molecular  mechanism with a cooperative behavior can  still rely on  interacting bi-stable elements but    the local spin interactions would have to be replaced by  the global  interactions  transmitted  through sufficiently stiff backbones. 
 
To  show how such bio-mimetically inspired design idea can be implemented in engineering systems, we study in this paper the simplest mechanical example: a parallel bundle of snap springs loaded by a force.  This toy model mimics  the power stroke mechanism  in acto-myosin contractile apparatus allowing a  sub-micron muscle sarcomere  to recover an abruptly applied slack at a time scale of \( 1\, \)ms despite the exposure to a thermal bath and without any use of metabolic resources (ATP) \citep{Caruel:2016kw}. 

The abrupt transitions at finite temperatures become possible when the free energy (rather than elastic energy) is  nonconvex,  and it is known in statistical mechanics that  this can take place in macroscopic systems  due to the presence of  long-range interactions \citep{Lebowitz:1966dj}. For us, of particular interest in this respect are the  zero dimensional systems with  infinitely long range (mean-field)  interactions \citep{Kometani_1975,Desai_1978}.  In the context of  fast force recovery in skeletal muscles, a mean-field  model  was first proposed in the   seminal paper by Huxley and Simmons (HS)~\citep{Huxley_1971,Caruel:2016kw},  where the authors considered a parallel bundle of bi-stable elements subjected to white noise. However,  since   the study of HS  was performed in a  hard device  (length clamp) prohibiting collective effects,  the cooperative behavior was not found. While the HS  model  has been since generalized \citep{Hill:1974ks,Hill:1976gf,Huxley_1996,Piazzesi_1995,Linari_2010a,Smith_2008,Caruel:2016kw}  and  reformulated in the context of cell adhesion \citep{Bell_1978} and hair cell gating \citep{Howard:1988uu}, the issue of hypersensitivity of the mechanical response and the possibility of   abrupt transitions at finite temperature in response to incremental loading have not been addressed.

In the present paper, we  take advantage of the fact that the   behavior of systems with long-range interactions is ensemble dependent \citep{Barre:2002ck} 
and  study  the distinctly different behavior of the HS model in the soft  device conditions (load clamp).
This seemingly  innocent generalization  of the original HS setting   brings  with it the desired mean-field elastic interactions that ensure that a parallel bundle of bi-stable units loaded by a force  can  maintain a synchronized state  in the presence of thermal fluctuations.

Moreover we show that such mechanical system in a soft device can be mapped into a ferromagnetic Ising model with mean-field interactions, which is known to  exhibits not only metastability abut also  critical behavior at the particular values of parameters.

To substantiate this analogy, we  present a systematic study of  mechanical and thermal properties for a finite size HS model  with a soft device loading. We show that for this mechanical system one can identify a critical temperature where both mechanical and thermal susceptibilities diverge.
 Below this temperature,  the system is macroscopically bi-stable, exhibiting synchronized fluctuations between two coherent states.
 Given that the transitions between the two states  are abrupt,  such  fluctuations can be viewed as a representation of domain structure in  time (rather than space).
 Above the critical temperature, the non-equilibrium free energy of the system  is convex with a minimum describing uncorrelated fluctuations and corresponding to ``non-snap-spring'' behavior.
 We show that the existence of a critical point can be already felt at finite \( N \), and  that the degree of synchronization can be  tuned by varying the number of elements.
In addition to equilibrium behavior, we  study  the kinetic  response of the system, and show that sharp equilibrium transitions in quasi-static conditions  give rise to hysterestic behavior at finite loading rates.
The survival of the jumps inside the hysteresis loop emphasizes the  robustness of the discontinuous behavior. 

The paper is organized as follows. In Sec.~\ref{sec:single_element}, we analyze the behavior  of a  single element bi-stable system and conclude that in this case the abrupt response  is  always compromised by finite temperature. In Sec.~\ref{sec:parallel_bundle_of_n_elements}, we   study the finite temperature transition between synchronized and non-synchronized responses of a finite size cluster of interacting bi-stable elements. The properties of the critical point are detailed in Sec.~\ref{sec:critical_behavior}.
The kinetic behavior of cluster of snap-spring is studied in Sec.~\ref{sec:dynamic_loading}. Our results are summarized  in Sec.~\ref{sec:conclusions}. In   the three Appendices A, B and C we briefly discuss the equilibrium susceptibility,  the thermal behavior of the system and the limits of bi-stability.

In a companion paper \citep{Caruel:2016kw} we show that although the hard device version of the HS model does not exhibit a cooperative snap-through behavior, its mechanical response is characterized by an intriguing negative stiffness.

\section{Single element} 
\label{sec:single_element}

\begin{figure}
	\centering
	\includegraphics[]{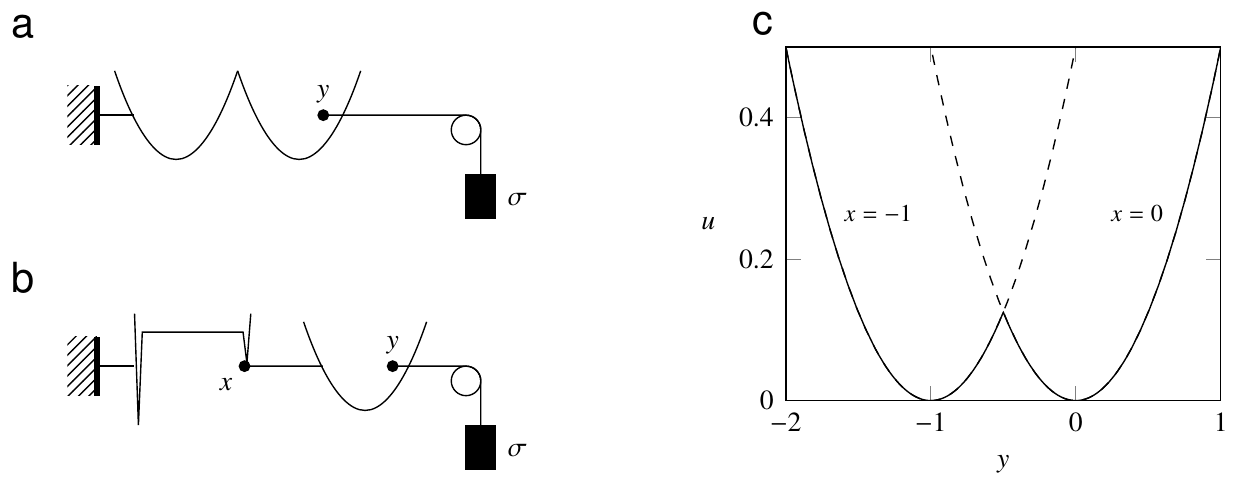}
	\caption{
	Model of a single mechanical switch. (a) Snap-spring model; (b) Digital switch (spin) model. (c) Internal energy landscape of the device as function of its total elongation \( y \) ; solid line, mechanical switch; dashed line, digital switch. Here \( v_0=0 \).
	}
	\label{fig:SingleSwitchModel}
\end{figure}
Consider   a bistable spring with  one degree of freedom $y$ shown in Fig.~\ref{fig:SingleSwitchModel}(a). Its  double-well energy, which for analytical transparency  we choose to be bi-quadratic, can be written as
\begin{equation*}
	u(y)=(1/2)\left(\left|y+1/2\right|-1/2\right)^2,
\end{equation*}
see the solid line in Fig.~\ref{fig:SingleSwitchModel}(c). The system has two stable states located at \( y=-1 \) and \( y=0 \), that are separated by an energy barrier located at \( y=-1/2 \).

Suppose that this device is submitted to an external force \( \sigma \), so that the total energy of the system is
\(
	w(y)=u(y)-\sigma y
\).
In equilibrium, a metastable configuration must satisfy
\[ \left(y+1/2\right)\left[1-1/(2\left|y+1/2\right|)\right]=\sigma,\]
which means, in particular,  that at \( \sigma =0 \), the system following the global minimum path will undergo a finite jump in $y$. 

At a finite temperature $T$, the same  system is  described by the probability distribution 
\[
	\rho(y,\beta) = Z^{-1}\exp\left[-\beta\, w(y;\sigma)\right],
\]
where \( \beta = 1/(k_{b}T) \), with \( k_{b} \) being Boltzmann's constant.
From the partition function \footnote{Here we use the complementary error function 
\( \mathrm{erfc}(x) = \frac{2}{\sqrt{\pi}}\int_x^{\infty}\exp\left[-t^{2}\right]dt. \)
}
\begin{multline*}
	Z(\sigma, \beta) = \int_{-\infty}^{\infty} \exp[-\beta\, w(y;\sigma] dy = \sqrt{\frac{\pi}{2\beta}}
	\left\{
	\exp\left[-\beta\left(\sigma-\frac{\sigma^{2}}{2}\right)\right]\mathrm{erfc}\left[\sqrt{\frac{\beta}{2}}\left(\sigma-\frac{1}{2}\right)\right]\right.\\
	\left.
	+
	\exp\left[-\beta\left(-\frac{\sigma^{2}}{2}\right)\right]\mathrm{erfc}\left[\sqrt{\frac{\beta}{2}}\left(\sigma+\frac{1}{2}\right)\right]
	\right\},
\end{multline*}
we obtain the equilibrium free energy \[ g(\sigma,\beta) = -\beta^{-1}\log\left[Z(\sigma,\beta)\right]. \]
The constitutive relation linking the control parameter $\sigma$ and the average elongation \( \mean{y} \)  takes the form 
\begin{equation*}
	\mean{y}(\sigma;\beta) = -\frac{\partial}{\partial \sigma}g(\sigma,\beta) = \sigma - \left\{
	1 + \exp\left[\beta\,\sigma\right] 
	\frac{
	\mathrm{erfc}\left[-\sqrt{\beta/2}\left(\sigma+1/2\right)\right]
	}
	{
	\mathrm{erfc}\left[\sqrt{\beta/2}\left(\sigma-1/2\right)\right]
	}
	\right\}^{-1}.
\end{equation*}

\begin{figure}
	\centering
	\includegraphics[]{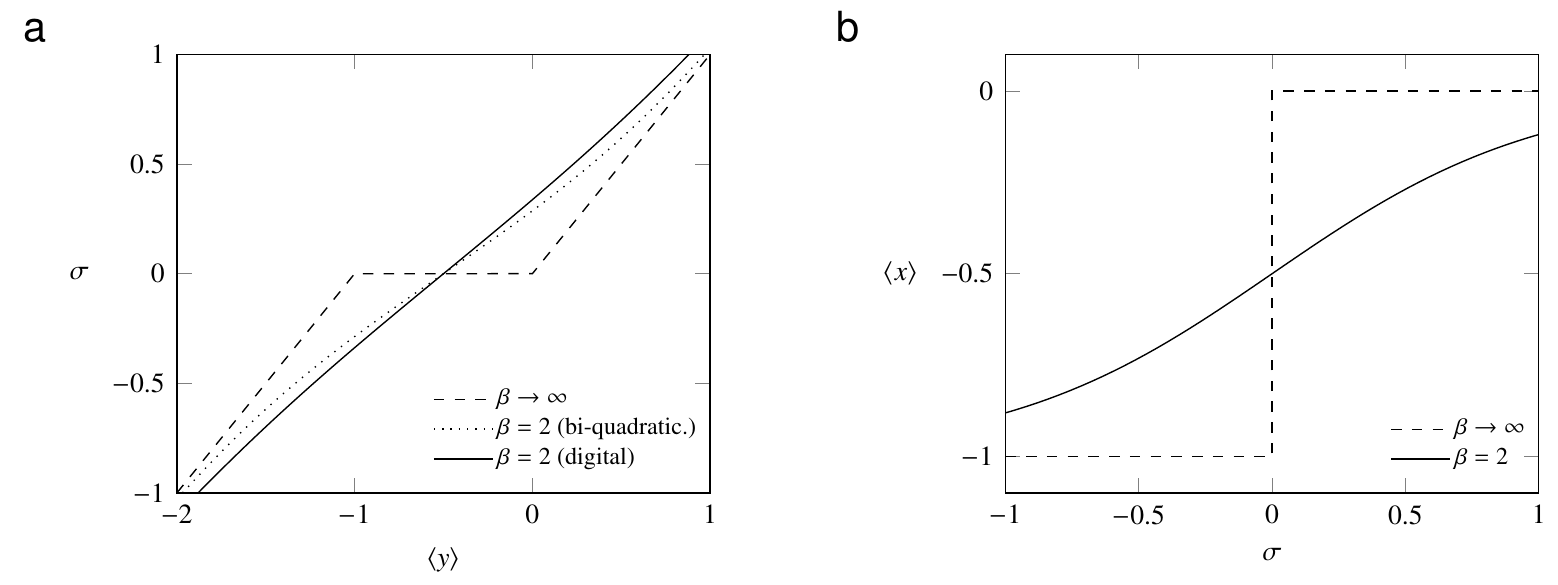}
	\caption{
 		Single switching element at finite temperature. 
		(a) Average elongation as function of the applied load; At zero temperature (\( \beta\to\infty \), dashed line), the response is the same with both the two models. At finite temperature (\( \beta=2 \)) the bi-quadratic (dotted) and the digital switch (solid) show similar smoothened responses.
		(b) Average internal configuration of the digital spin model at zero temperature (dashed line) and at finite temperature (solid line). Results are shown for \( v_0=0 \).
	}
	\label{fig:single_switch}
\end{figure}

The resulting tension elongation relation is illustrated in Fig.~\ref{fig:single_switch}(a). 
While at zero temperature (dashed line), the desired switch-like character of the response is clearly visible with an abrupt transition taking place at \( \sigma=0 \), at any finite temperature (dotted and solid lines), the response is  smooth.
 
To show that this result   is independent on the detailed structure of the snap spring element, we now consider another representation of a bi-stable device \citep{Huxley_1971}. We introduce a spin  variable \( x \),  and link it through a linear spring of  stiffness \( \kappa_0 \) with a continuous variable \( y \). The energy associated with the spin variable \( x \) is  
\begin{equation*}
	u_{0}(x) = 
	\begin{cases}
		v_0	& \text{if } x = 0\\
		0		& \text{if } x = -a\\
		\infty	&	\text{otherwize}
	\end{cases}
\end{equation*}
Here \( a \) is the  ``reference'' length  of the spin element, and   \( v_{0} \) is the intrinsic energy bias between the two states; see Fig.~\ref{fig:SingleSwitchModel}(b).
We take \( a \) as our characteristic distance and \( \kappa_0 a^{2} \) as the characteristic energy. Hence the non-dimensional energy of the spin element  can be written as 
$ u(x,y)=(1+x)v_{0}+(y-x)^{2}/2, $
 where now \( y \) represents the continuous total elongation while \( x \) takes the discrete values \( -1 \) or \( 0 \).
At a given elongation, our element has two states \( v_0 + y^{2}/2 \) and  \( (y+1)^{2}/2 \),  which are represented by dashed lines in Fig.~\ref{fig:SingleSwitchModel}(c). These two states have the same energy at \[ y=y_{0}\equiv v_0-1/2. \]

Using dimensionless variables we can write the total energy of such system exposed to the force $\sigma$   as
\begin{equation}
	\label{eq:w_single_element}
	w(x,y;\sigma)=(1+x)\,v_0 + (1/2)\,(y-x)^{2}-\sigma y.
\end{equation}
At zero temperature, we again obtain an abrupt transition.
Indeed, if we minimize out the   elongation \( y \), the energy \eqref{eq:w_single_element} takes the form
\[
 \hat{w}(x;\sigma) = -\frac{1}{2}\sigma^{2}-\sigma x +\left(1+x\right)v_0.
\]
The global minimum of this energy correspond to   \( x=-1 \) for \( \sigma<\sigma_{0} \), and   \( x=0 \) for \( \sigma>\sigma_0 \), with the transition occurring at \[ \sigma=\sigma_{0} \equiv v_0. \]
Therefore, for \( v_0=0 \),  the digital switch element [Fig.~\ref{fig:SingleSwitchModel}(b)] and the snap spring element [Fig.~\ref{fig:SingleSwitchModel}(a)] behave similarly at zero temperature; see Fig.~\ref{fig:single_switch}[(a), dashed line].

At finite temperature \( T \), the spin device is  described by the probability distribution
\[
	\rho(x,y;\sigma,\beta) = Z^{-1}\exp\left[-\beta\,w(x,y;\sigma)\right],
\]
where the nondimensional measure of the thermal fluctuations is \[ \beta^{-1} = (k_{b}T)/(\kappa_{0}a^{2}). \] The partition function  is given by
\begin{equation}
	\label{eq:single_switch_partition_function}
	Z(\sigma,\beta) = \int_{-\infty}^{\infty}  \sum_x \exp\left[-\beta\,w(x,y;\sigma)\right]dy\\
	=\sqrt{\frac{2\pi}{\beta}}
	\exp\left[
	-\frac{\beta}{2}\left(-\sigma^{2}+\sigma+\sigma_{0}\right)
	\right]
	2\cosh\left[\frac{\beta}{2}\,(\sigma-\sigma_0)\right].
\end{equation}
The free energy then takes the form,
\begin{equation*}
	g(\sigma,\beta)=
	-\frac{\sigma^{2}}{2} + \frac{\sigma-\sigma_0}{2} + \sigma_0
	-\frac{1}{\beta}\log\left\{2\cosh\left[\frac{\beta}{2}\,(\sigma-\sigma_0)\right]\right\}-\frac{1}{2\,\beta}\log\left[\frac{2\pi}{\beta}\right],
\end{equation*}
while the relation between the control parameter $\sigma$ and the average elongation \( \mean{y} \) reads
\begin{equation*}
	\mean{y}(\sigma,\beta) = -\frac{\partial g}{\partial\sigma} =  \sigma-\frac{1}{2}+\frac{1}{2}\tanh\left[\frac{\beta}{2}\left(\sigma-\sigma_0\right)\right].
\end{equation*}
As in the case of the snap spring device at finite temperature, the discontinuity at \( \sigma=0 \) disappears. Since the difference between the two systems is only of quantitative nature, see Fig.~\ref{fig:single_switch}[(a), dotted and solid lines], 
 we abandon in what follows the soft spin description and focus exclusively on the digital switch model which is analytically more transparent.

In addition to the total elongation, we have access to the average internal configuration of the spin element, which can be written as
\begin{equation*}
	 \mean{x}(\sigma,\beta) = \sum_{x} \int x\,\rho(x,y;\sigma,\beta)dy = \mean{y}(\sigma,\beta) - \sigma =-\frac{1}{2}\left\{1-\tanh\left[\frac{\beta}{2}\left(\sigma-\sigma_0\right)\right]\right\}.
\end{equation*}
The dependence of $\mean{x}$ on the applied force is shown in Fig.~\ref{fig:single_switch}(b), for the case \( v_0=0 \). At zero temperature we observe the predicted discontinuity at \( \sigma=0 \). At finite temperature the response loses its switch-like behavior: it is destroyed by thermal fluctuations because they allow the system to visit both spin states at fixed \( \sigma \).

In  \ref{sec:stiffness},  we complement our analysis of a single element system by studying the behavior of its equilibrium \emph{susceptibility} and a closely related elastic \emph{stiffness}.
In  \ref{sec:thermal_behavior}, we discuss its  \emph{thermal} properties  particularly focusing on the parametric dependence of the  entropy and the specific heat.


\section{Parallel bundle of \( N \) elements} 
\label{sec:parallel_bundle_of_n_elements}

\begin{figure}
	\centering
	\includegraphics[]{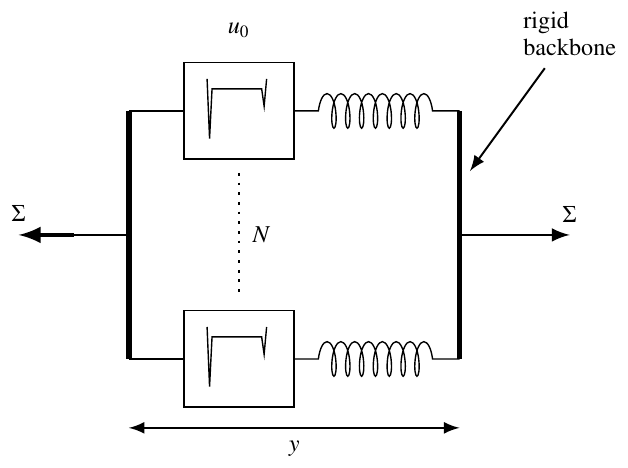}
	\caption{
	Mechanical model of a cluster of \( N \) digital switches in parallel submitted to an external force \( \Sigma \).
	}
	\label{fig:ClusterModel}
\end{figure}
To overcome the virtual 
``desintegration''  of the switching device at finite temperature, we now consider \( N \) elements attached in parallel between two rigid backbones; see Fig.~\ref{fig:ClusterModel}. The internal energy 
\begin{equation}\label{eq:HS_v}
	v(\boldsymbol{x};z) = \frac{1}{N} \sum_{i=1}^{N} \left[\left(1+x_{i}\right)v_0 + \frac{1}{2}(y-x_{i} )^{2}\right],
\end{equation}
depends only on two non-dimensional parameters, \( v_{0} \) and \( N \). Here
 \( \boldsymbol{x} \) denotes the vector \( (x_{1},\dots,x_{N}) \). In a soft device  setting (force clamp),  we prescribe the total tension \( \Sigma \), and the energy per element can be written as \( w = v - \sigma y \), where \( \sigma=\Sigma/N \). Note that a switch  in a  single unit has to be compensated by the changes in the other elements  to ensure the force balance. This  mechanical feedback can be characterized as   a mean-field interaction.



At finite temperature, the probability density associated with a microstate \( \boldsymbol{x} \) is given by
\begin{equation*}
	\rho(\boldsymbol{x},y;\sigma,\beta) =
	 Z(\sigma,\beta)^{-1}\exp\left[-\beta N w(\boldsymbol{x},y;\sigma)\right],
\end{equation*}
where  
\begin{equation*}
	Z(\sigma,\beta)= \int\sum_{\boldsymbol{x}\in\{0,1\}^{N}}\exp\left\{-\beta N\left[ v(\boldsymbol{x};y)-\sigma y\right] \right\}dy.
\end{equation*}

Due to the parallel connection, the system is invariant with respect to permutation of the variables  \( x_{i} \). Therefore the sum over the microstates can be transformed into a sum over  
\[
	p=-\frac{1}{N}\sum_{i=1}^{N}x_{i},
\]
which represents the fraction of elements  in the ``short'' phase.
We thus obtain
\begin{equation}
	\label{eq:z}
	Z\of{\sigma,\beta} = \int\sum\limits_{Np=0}^{N}
	\binom{N}{Np}\exp\left\{-\beta N\left[ \bar{v}(p,y)-\sigma y \right] \right\}
	dy,
\end{equation}
where \( \binom{N}{Np}=\frac{N!}{(Np)!(N-Np)!} \) is the degeneracy at a given value of \( p \), and
\begin{equation}
	\label{eq:hat_v}
	\bar{v}(p;y,\beta) = p\frac{1}{2}(y+1)^{2} + (1-p)\left[\frac{1}{2}y^{2}+v_{0}\right],
\end{equation}
is the internal energy of a metastable state at fixed \( (p,y) \).
The partition function \eqref{eq:z} can also be put in the form
\begin{equation}
	\label{eq:z_marginal_distribution}
		Z(\sigma,\beta) = \int\sum_{p}\exp\left[-\beta N\, \bar{g}(p,y;\sigma,\beta)\right]dy
\end{equation}
where \( \bar{g} \) is the marginal free energy at fixed \( (p,y) \) defined by
\begin{equation}\label{eq:g_y_bar}
	\overline{g}(p,y;\sigma,\beta)  =  \bar{v}(p,y)-\sigma y- \frac{1}{\beta}s(p).
\end{equation}
Here \( \bar{v} \) is given by \eqref{eq:hat_v}, and the entropy is 
\[
	s(p) =  \frac{1}{N}\log \binom{N}{Np}.
\]
By integrating over $y$ in \eqref{eq:z_marginal_distribution}, we obtain another  marginal free energy  
 \begin{equation}
 	\label{eq:g_bar_p}
 	\bar{g}(p;\sigma,\beta) = \bar{w}(p,\sigma)  - \frac{1}{\beta} s(p)-\frac{1}{2\beta N}\ln\biggl(\frac{2\pi}{\beta N}\biggr),
 \end{equation}
 where,
 \begin{equation}
 	\label{eq:hat_w}
 	\bar{w}(p;\sigma) = \sigma_0-\frac{1}{2}\sigma^{2} + p(\sigma-\sigma_0) + \frac{1}{2}p(1-p),
 \end{equation}
 is the mechanical energy of a metastable state characterized by \( p \); see \cite{Caruel:2015im}. We recall here that in our dimensionless variables \( \sigma_0=v_0 \).
 The presence of the term \( p(1-p) \) in Eq.~\eqref{eq:hat_w} can be viewed as the signature of interactions providing a mechanical coupling between the elements. To obtain the equilibrium value of the elongation, we minimize \eqref{eq:g_y_bar} with respect to \( y \), and find  \( \bar{y} = \sigma-p \).

\begin{figure}[t]
	\centering
	\includegraphics[]{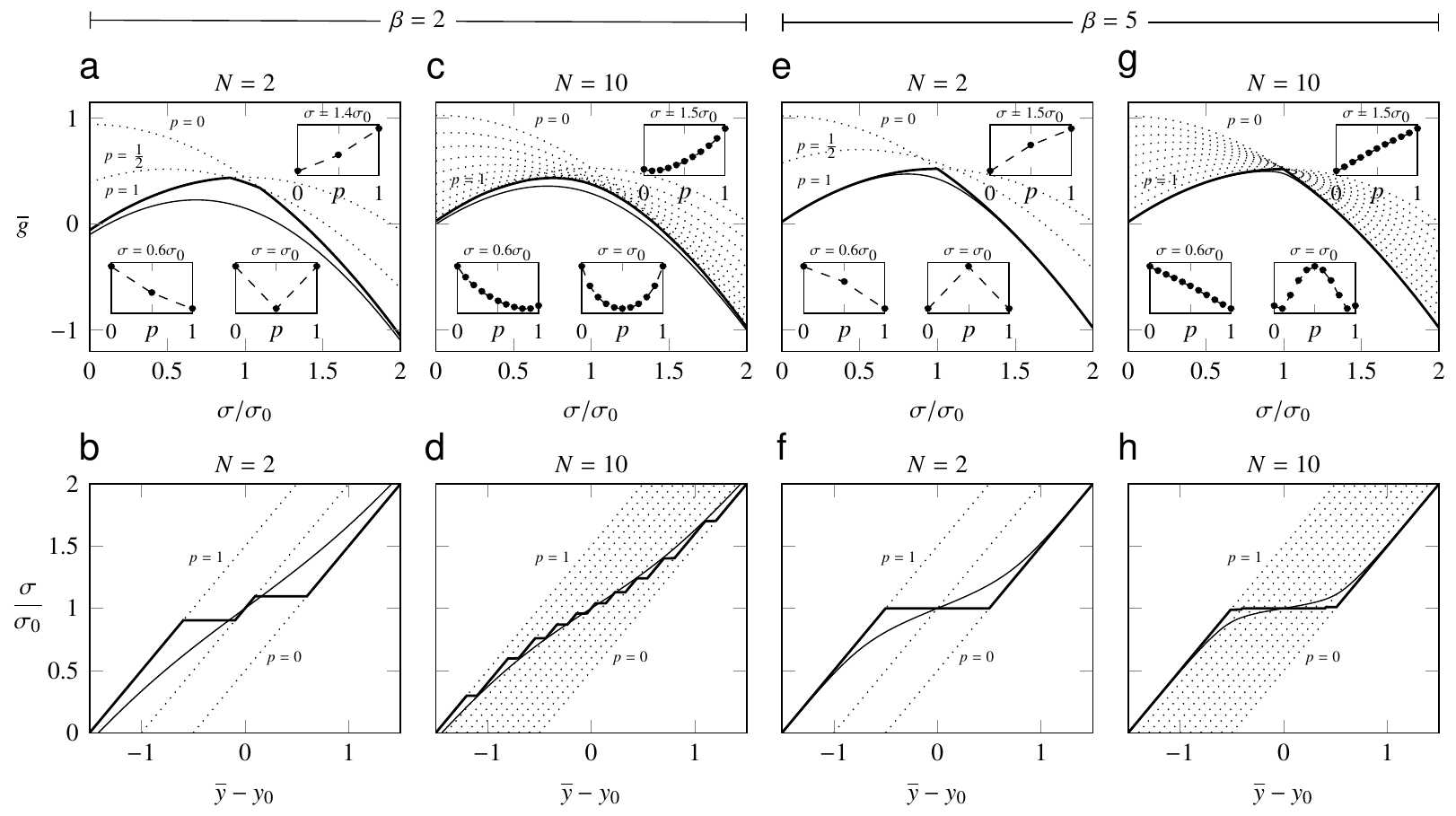}
	\caption{
	Marginal free energy  and tension-elongation of the HS model in a soft device at \( \beta=2 \) [(a) to (d)] and \( \beta=5 \) [(e) to (h)] for \( N=2 \) [(a), (b) and (e), (f)] and for \( N=10 \) [(c), (d) and (g), (h)].
First row, marginal free energy levels \( \bar{g} \) corresponding to different values of \( p \) [dotted lines, see Eq.~\eqref{eq:g_bar_p}], global minimum (bold line). Inserts show the free energy profiles for the selected values of the applied force. Second row, tension-elongation relations corresponding to each case of the first row with same line styles. In each plot, the thin solid line represents the behavior in thermal equilibrium.
	}
	\label{fig:MarginalFreeEnergyFiniteN}
\end{figure}



To illustrate these  formulas, we note that in the case \( N=2 \) [see Fig.~\ref{fig:MarginalFreeEnergyFiniteN} (a), (b), (e) and (f)], the marginal free energy can only take one of the following values,
\begin{align}
	\overline{g}(0;\sigma,\beta) &= -\frac{1}{2}\sigma^{2}+v_{0} + C(\beta)\nonumber,\\
	\overline{g}(1/2;\sigma,\beta) &= -\frac{1}{2}\sigma^{2} + \frac{1}{2}\sigma + \frac{1}{8} + \frac{1}{2}v_{0} + -\frac{1}{2\beta}\ln 2 + C(\beta)\label{eq:hs_g_N=2},\\
	\overline{g}(1;\sigma,\beta) & = -\frac{1}{2}\sigma^{2}+ \sigma +C(\beta)\nonumber,
\end{align}
with \( C(\beta) = -1/(4\beta) \ln[\pi/\beta] \).  In   Fig.~\ref{fig:MarginalFreeEnergyFiniteN} we present the energy profiles and the corresponding tension-elongation relations for systems with \( N=2 , 10 \) at \( \beta=2 \)
, see Fig.~\ref{fig:MarginalFreeEnergyFiniteN}[(a)-(d)]; and at \( \beta=5 \); see Fig.~\ref{fig:MarginalFreeEnergyFiniteN}[(e)-(h)]. For each value of \( \sigma \), the metastable state with the lowest free energy gives a point on the  global minimum path shown by the bold line.  Note that above a certain temperature  the free energy is a convex function of \( p \) (see inserts in Fig.~\ref{fig:MarginalFreeEnergyFiniteN}[(a) and (c)]), which implies that  the global minimum path goes through all mixed configuration with \( 0\leq p\leq 1 \). Below this temperature, the free energy is not convex (see inserts in Fig.~\ref{fig:MarginalFreeEnergyFiniteN}[(e) and (g)]) and the mixed states may have higher energy than the ordered states. 
As a result, the system following (as \( \sigma \) increases or decreases)  the global minimum path,  will experience an abrupt jump between the almost homogeneous metastable configurations with \( p \approx 0,1 \), while the mixed states are simply swept through. The resulting tension elongation curve will exhibit a plateau whose length indicates the size of the jump, see Fig.~\ref{fig:MarginalFreeEnergyFiniteN}[(f) and (h)].

\begin{figure}
	\centering
	\includegraphics[]{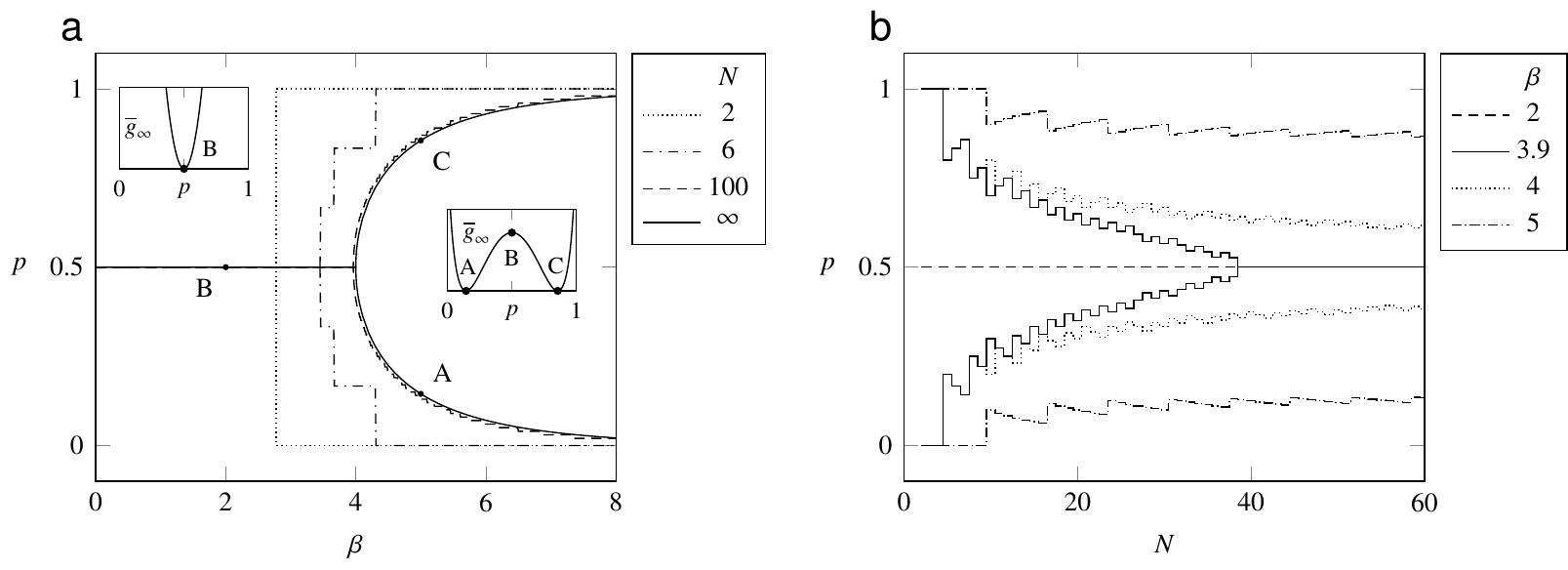}
	\caption{
	Bifurcation diagrams  at
	\( \sigma=\sigma_0 \),
	obtained by fixing the number of units and changing the temperature (a) and by fixing the temperature and changing the number of units (b). Each curve shows the local minima of the marginal free energy \eqref{eq:g_bar_p}. The inserts in (a) show typical marginal free-energy profiles corresponding to sub- and supercritical temperatures in the thermodynamic limit.
	}
	\label{fig:Bifurcations}
\end{figure}

Even though the global minimization of the marginal energy does not describe the equilibrium behavior of the system, the above observations suggest  that there is an order-disorder transition in the system (with pure states dominating at low temperatures, and mixed state dominating at high temperatures).
For finite \( N \), the parameters of the transition can be estimated by computing the discrete derivatives of the marginal free energy \eqref{eq:g_bar_p}, which will converge to the derivatives of \( \bar{g} \) in the thermodynamic limit. Thus, if we define the finite difference  \( \Delta^{2}\bar{g} = \bar{g}(p+1/N) + \bar{g}(p-1/N) -2\bar{g}(p) \) we  obtain, for \( N>2 \)
\begin{equation*}
	N^{2}\Delta^{2}\bar{g}(p;\sigma,\beta) = -1 +  \frac{N}{\beta}\log\left[1+\frac{N+1}{N^{2}p(1-p)}\right].
\end{equation*} 
This expression is positive for \( \beta<\beta_{c} \), with
\begin{equation*}
	\beta_{c} = N\log\left[1+4\frac{N+1}{N^{2}}\right],
\end{equation*}
and is negative otherwise. Hence, for \( \beta<\beta_c \), the system has a single value of \( p \) minimizing \eqref{eq:g_bar_p}, while for \( \beta>\beta_c \) the same system is bistable. Similarly for a given value of \( \beta \), we can find  a critical size \( N_{c} \) below which the system  becomes bistable.
The positions of the minima of the marginal free energy \eqref{eq:g_bar_p} are represented in Fig.~\ref{fig:Bifurcations} as functions of the temperature [see (a)] and of the number of elements [see (b)]  at the fixed  value of the external force \( \sigma=\sigma_0 \).

Below the critical temperature (\( \beta>\beta_{c} \)), or below the critical system size (\( N<N_{c} \)), the marginal free energy has two minima corresponding to highly synchronized configurations. 
Notice that at this particular loading (\( \sigma=\sigma_0 \) and subcritical temperature \( \beta>\beta_{c} \)), the two minima of the free energy have the same depth (see the inserts of Fig.~\ref{fig:MarginalFreeEnergyFiniteN}) and the free energy itself is symmetric independently of the choice of \( v_0 \).
The size of the jump observed along the global minimum path corresponds in Fig.~\ref{fig:Bifurcations}(a) to the distance between the two branches of the pitchfork bifurcation.
 
When \( N\to\infty \) the values of the critical temperature converge to \( \beta_{c} = 4 \), and Fig.~\ref{fig:Bifurcations}(a) shows that this estimate is already adequate at \( N=100 \) (see dashed line). Hence, it is possible to control the degree of synchronization by changing  the cluster's size when  \( N\sim 10 \). 
\begin{figure}
	\centering
	\includegraphics[]{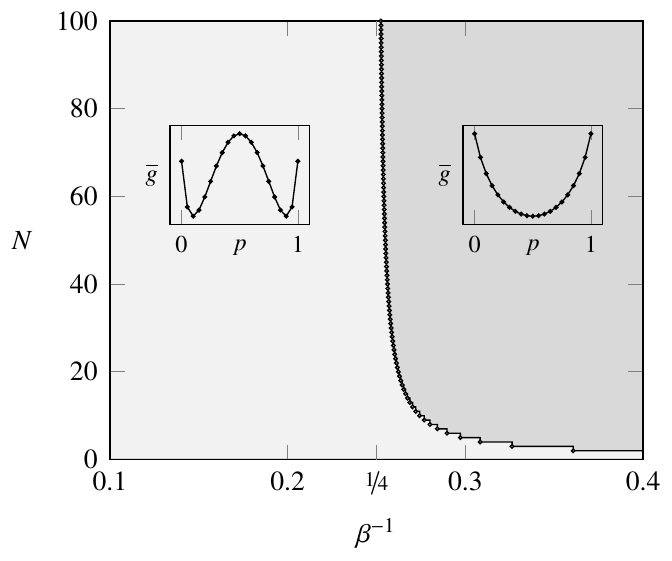}
	\caption{
	Phase diagram for finite \( N \), showing the domains where the system is cooperative (light gray) and noncooperative (dark gray). Inserts show typical marginal free energy profiles at \( \sigma=\sigma_0 \) in both phases.
	}
	\label{fig:PhaseDiagram}
\end{figure}
Above \( N=100 \), the degree of synchronization does not depend anymore on the number of elements. These  observations are summarized in the phase diagram shown in Fig.~\ref{fig:PhaseDiagram}.



The formulas for the marginal free energy  become more transparent in the thermodynamic limit \( N\to\infty \)  when
we can use the Stirling formula \( \log(n!) \approx n\log(n)-n \) and write the marginal free energy \eqref{eq:g_y_bar} as
\begin{equation}
	\label{eq:g_bar_y_infinity}
	\bar{g}_{\infty}(p,y;\sigma,\beta) = \bar{v}(y,p) - \sigma y - \left(1/\beta\right)s_{\infty}(p).
\end{equation}
Here 
$
	s_{\infty}(p) = -\left[p\log(p) + (1-p)\log(1-p)\right],
$
is the ideal mixing entropy.
\begin{figure}
	\centering
		\includegraphics[]{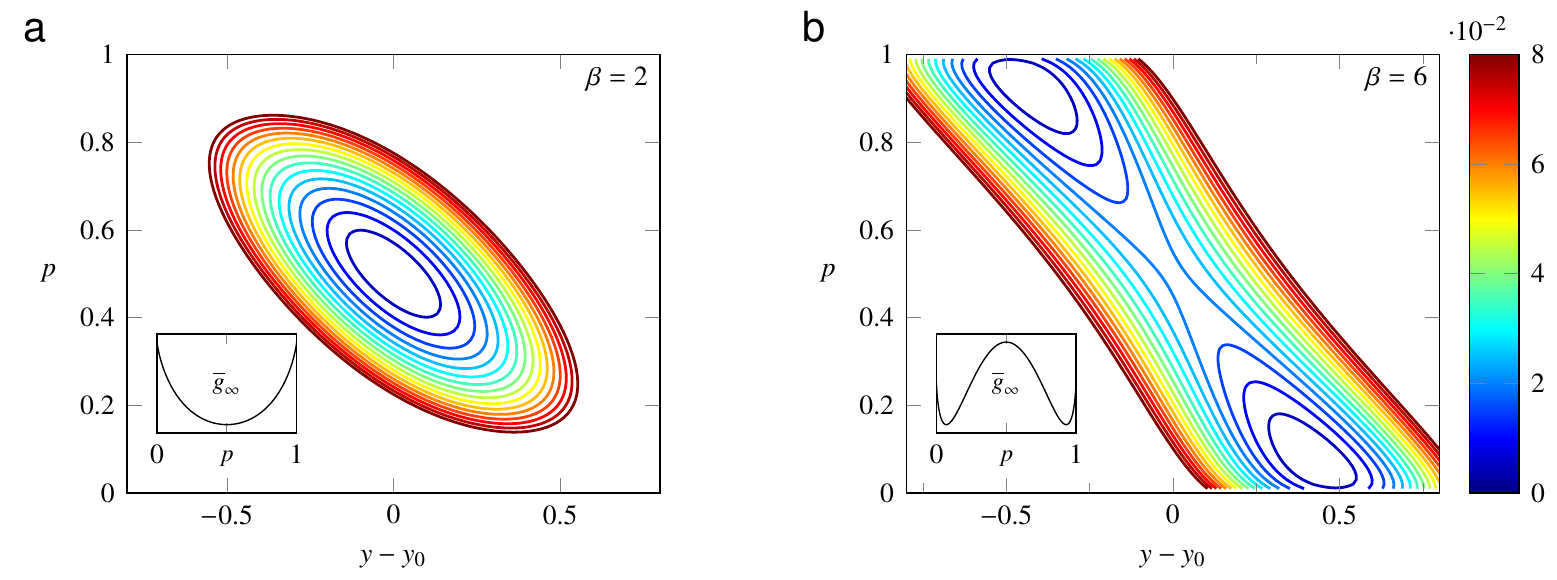}
	\caption{Marginal free energy \( \bar{g}_{\infty} \)  at \( \sigma=\sigma_{0} \) in the thermodynamic limit \( N\to\infty \). (a) \( \beta=2 \), (b) \( \beta=6 \). The inserts show the energy profile of \( \bar{g}_{\infty} \) after elimination of \( y \). Energy minima are set to 0 for comparison and the contours associated to energy higher than 0.08 are not shown for the sake of clarity.
 	}
	\label{fig:g_level_set}
\end{figure}
These formulas are illustrated in Fig.~\ref{fig:g_level_set}, where the level sets of the marginal free energy \eqref{eq:g_bar_y_infinity} are represented at \( \sigma=\sigma_0 \) for \( \beta=2 \) [see (a)] and \( \beta=6 \); see (b).
We again see that the free energy is convex at large temperatures (a), and non-convex---with two macroscopic synchronized minima---at low temperatures (b). 
The critical temperature for this  transition can be computed explicitly from the determinant of the Hessian  matrix 
\begin{equation*}
	H(p,\beta) = 
	\begin{pmatrix}
		1&1\\
		1&\left[\beta p\left(1-p\right)\right]^{-1}
	\end{pmatrix},
\end{equation*}
which is positive if \( \beta < \min_{p\in[0,1]}[p(1-p)]^{-1}=\beta_{c}=4 \). The ensuing pitchfork bifurcation is represented by the solid line in Fig.~\ref{fig:Bifurcations}(a).

The physical origin of the   transition at $\beta_{c}$ becomes more transparent   if we introduce the new variable \( \xi(p,y;\sigma) = y+p-\sigma  \), and rewrite the free energy \( \bar{g}_{\infty} \) as a function of \( \xi \) and \( p \)  
\begin{equation}
	\label{eq:g_yp_with_xi}
	\bar{g}_{\infty}(\xi,p;\sigma,\beta) = \frac{1}{2}\xi^{2} + \bar{w}(p;\sigma) - \frac{1}{\beta} s_{\infty}(p),
\end{equation}
where \( \bar{w} \) is given by Eq.~\eqref{eq:hat_w}.
We see that in Eq.~\eqref{eq:g_yp_with_xi} the function \( \bar{g}_{\infty} \) is always convex in \( \xi \). If we minimize out \( \xi \), we obtain the  one dimensional marginal free energy  
\begin{equation}
	\label{eq:g_p_infinity}
		\bar{g}_{\infty}(p;\sigma,\beta) = \bar{w}(p,\sigma) - \frac{1}{\beta} s^\infty(p).
	\end{equation}
Here the mechanical energy of the system \( \bar{w} \) is always concave in \( p \), implying that its  minimum is reached at one of the two homogeneous states \( p=0 \) or \( p=1 \), while the  mixed states are endowed with higher mechanical energy due to the mean-field interaction (manifested  through the mixing energy  \( p(1-p)/2 \) in Eq.~\ref{eq:hat_w}).  Instead,  the entropy \( s(p) \) is always convex with a maximum at \( p=1/2 \).
Therefore   the convexity of the free energy is an outcome of the competition between the term \( p(1-p) \) and the term \( s_{\infty}(p)/\beta \), with the later becoming dominant  at low \( \beta \).

 \begin{figure}
	 \centering
 	\includegraphics[]{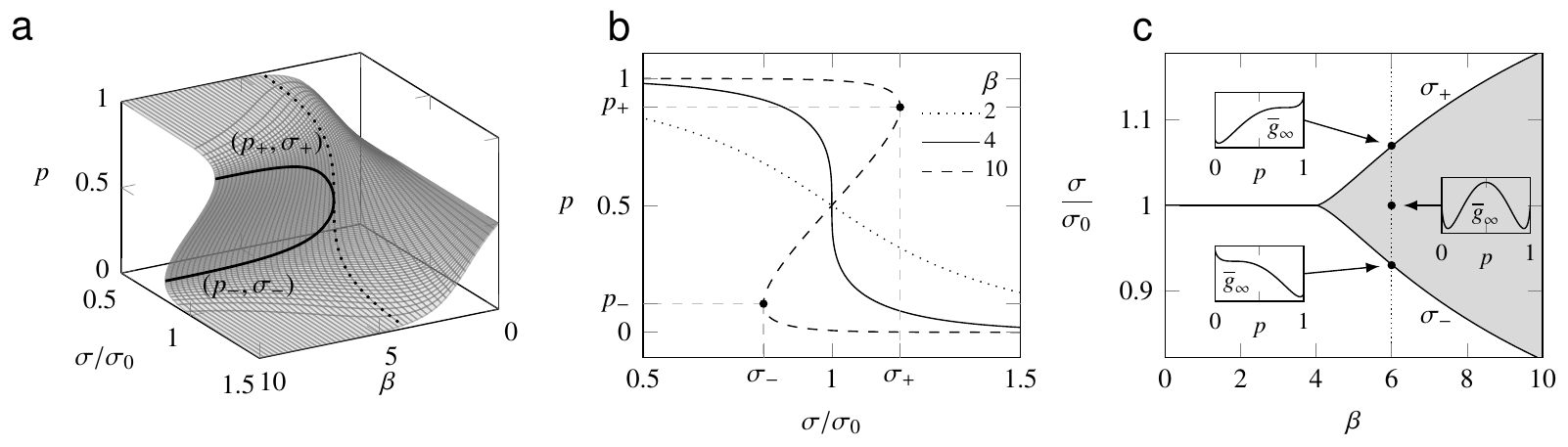}

	\caption{Bistability domain. (a) Representation of the critical points of the marginal free energy \( \bar{g}_{\infty} \) as function of temperature and loading. 
Dotted line, location of the minimum of the free energy \( \bar{g}_{\infty} \) at the critical point \( \beta=4 \); bold line, boundaries of the bistable domain.
(b) Location of the minima of \( \bar{g}_{\infty} \) for three different temperatures corresponding to three isotherms of (a).
(c) Evolution of the bistable domain (grey area) delimited by the critical lines \( \sigma_{\pm} \) as function of temperature. The inserts show the marginal free energy \( \bar{g}_{\infty} \) for \( \sigma=\sigma_{\pm} \) (left) and \( \sigma=\sigma_{0} \) at \( \beta=6 \).
}
	\label{fig:3d_bifurcation_p}
 \end{figure}

Energy profiles of \( \bar{g}_{\infty} \) at \( \sigma=\sigma_0 \) are represented as function of \( p \) in the inserts of Fig.~\ref{fig:g_level_set} for \( \beta=2 \) [see (a)] and \( \beta=6 \) [see (b)]. Above the critical point, the energy profile shows two metastable states that  are illustrated  in Fig.~\ref{fig:3d_bifurcation_p}.
Below the critical temperature, the domain of metastability spans the interval \( [\sigma_{-}(\beta),\sigma_{+}(\beta)] \) whose boundary (derived explicitly in the thermodynamic limit in \ref{sec:boundaries_of_the_bistable_domain}) corresponds to a first order phase transition, see Fig.~\ref{fig:3d_bifurcation_p} [(b) and (c)].
 \begin{figure}[tbp]
 	\centering
	\includegraphics{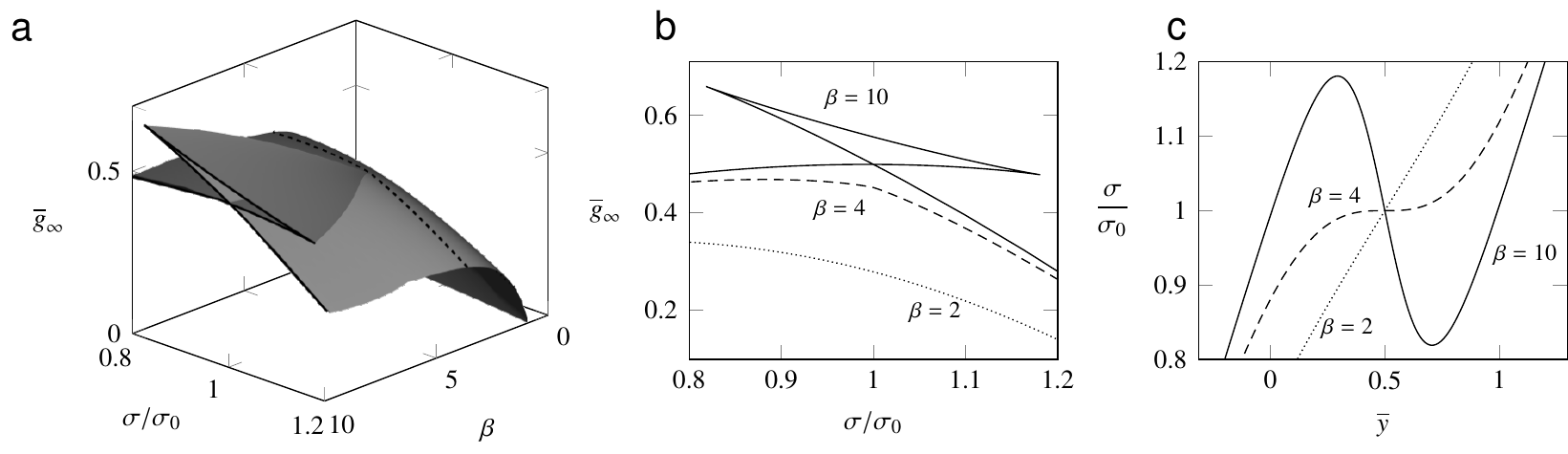}
 	\caption{
Mechanical properties of the metastable states. (a) Marginal free energy shown as function of temperature and tension. For \( \beta > 4 \) (marked with the dotted line), and within the interval \( [\sigma_-,\sigma_+] \), three energy levels coexist for the same tension. (b) Energy in the metastable states corresponding to three different temperatures (dotted, \( \beta=2 \); dashed, \( \beta=4 \); solid, \( \beta=10 \)). (c) Tension-elongation relations for the three same temperatures
 	}
 	\label{fig:3d_bifurcation_energy}
 \end{figure}
The energies of the metastable states are shown in Fig.~\ref{fig:3d_bifurcation_energy}[(a) and (b)]. For each of these states, the elongation can be recovered from the relation \( \bar{y}(p;\sigma,\beta) = \sigma - p \) obtained by adiabatic elimination of \( \xi \) in Eq.~\eqref{eq:g_yp_with_xi}. Examples of such tension-elongation relations  are shown in Fig.~\ref{fig:3d_bifurcation_energy}(c).



Using the marginal free energy \( \bar{g} \) given by Eq.~\eqref{eq:g_bar_p}, we can rewrite the partition function \eqref{eq:z_marginal_distribution} as 
\begin{equation}\label{eq:partition_function_with_g_of_p}
	Z(\sigma,\beta) = \sum_{p}\exp\left[-\beta N\,\bar{g}(p;\sigma,\beta)\right],
\end{equation}
and define the probability to find the equilibrium system in a particular metastable state 
\begin{equation}
	\label{eq:rho_bar}
	\bar{\rho}(p;\sigma,\beta) = Z^{-1}(\sigma,\beta)\exp\left[-\beta N\,\bar{g}(p;\sigma,\beta)\right].
\end{equation}
 The average equilibrium  configuration  is then characterized by
\[
	\mean{p}(\sigma,\beta) = \sum_{p} p\, \bar{\rho}(p;\sigma,\beta).
\]
In the thermodynamic limit, we can replace the sum in this expression by an integral over the interval \( [0,1] \) and obtain
$
	\mean{p}_{\infty}(\sigma,\beta) = p_{*}(\sigma,\beta),
$
where \( p_* \) is the minimum of the marginal free energy  \( \bar{g}_{\infty} \), see Eq.~\eqref{eq:g_p_infinity}. We have seen that the marginal free energy has two minima for \( \beta>\beta_{c} =4\). In the thermodynamic limit, the minimum with the lowest energy determines the equilibrium free energy and therefore, below the critical temperature the value of \( p_* \) jumps at \( \sigma=\sigma_0 \). 
 \begin{figure}[tbp]
 	\centering
 	\includegraphics[]{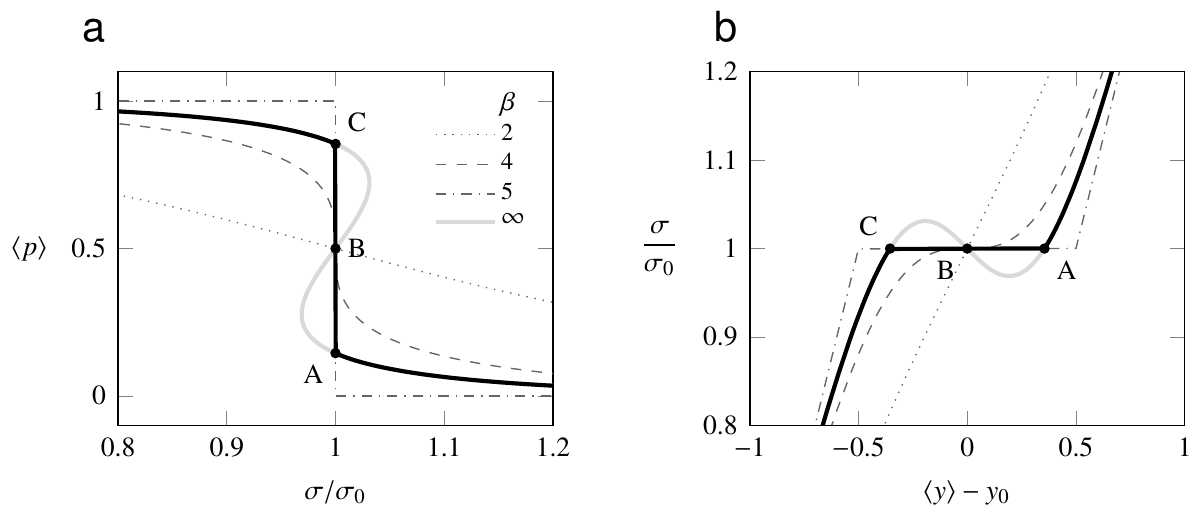}\\
	\includegraphics[]{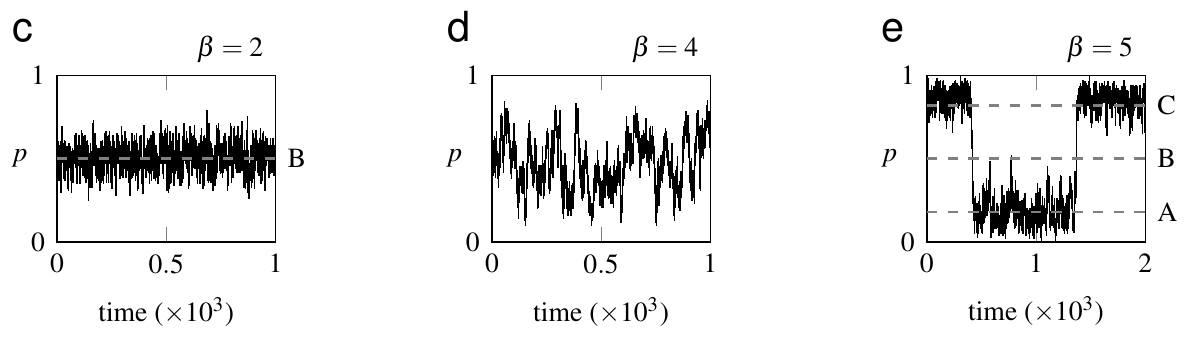}
 	\caption{
 	Average internal configuration (a) and average elongation (b) obtained in the thermodynamic limit for several temperatures. In the case \( \beta=5 \) (thick lines), we represent both the equilibrium curve (black)  and the metastable states corresponding to the critical points of the marginal free energy \eqref{eq:g_p_infinity} (gray). (c--e) Collective dynamics with \( N=100 \) in a soft device under constant force \( \sigma=\sigma_0 \)---ensuring \( \mean{p}=1/2 \)---for \( \beta =2 \) (c), \( \beta=4 \) (d) and \( \beta=5 \) (e). The letters A, B and C label the different metastable states in the case \( \beta=5 \).
 	}
 	\label{fig:Average_configuration}
 \end{figure}

We illustrate the dependence of the average configuration on  the loading in Fig.~\ref{fig:Average_configuration}(a). 
Above the critical temperature (\( \beta\leq 4 \)), the average configuration evolves smoothly between 1 and 0; while below it (\( \beta > 4 \)), we observe  an abrupt transition  at \( \sigma=\sigma_0 \); see the thick line in Fig.~\ref{fig:Average_configuration}(a).
In Fig.~\ref{fig:Average_configuration}[(c)--(e)], typical stochastic trajectories (with \( N=100 \)) of the order parameter \( p \) are represented for different temperatures, where the tension \( \sigma=\sigma_0 \) is fixed to ensure that \( \mean{p}=1/2 \).
One can see that in the  high temperature (paramagnetic) state [see Fig.~\ref{fig:Average_configuration}(c)] the elements  are in random conformations and the system fluctuate around the average value \( p=1/2 \) (labeled B), corresponding to the single minimum of the free free energy; see Fig.~\ref{fig:3d_bifurcation_p}[(a) and (b)].
Instead, in the  low temperature (ferromagnetic) state [see Fig.~\ref{fig:Average_configuration}(e)],
the average value \( p=1/2 \) corresponds to a local maximum of the free energy, which is an unstable state. As we illustrate in Fig.~\ref{fig:Average_configuration}(e), thermal equilibrium in this regime is ensured by coherent fluctuations of the system between two synchronized states (labeled A and C).

We recall that in classical ferromagnetism, the average value of the magnetization, say \( p=1/2 \), is ensured by the formation of spacial domain structure within the material. In our case of zero dimensional (mean-field) system, the spacial microstructure is replaced by a \emph{temporal microstructure} formed by  rare jumps between two coherent and long-living ordered configurations with \( p\approx 1 \) or \( p\approx 0 \).  
The time between the jumps is much larger than the characteristic time of the fluctuations near the metastable states. 
One can say that our switching device is endowed with a limited ``mechanical memory'' which is erased when the observation time (or loading time) exceeds the lifetime of the metastable states.

The mechanical  properties of the equilibrium system can be obtained from the knowledge of the partition function \eqref{eq:partition_function_with_g_of_p}. 
 The equilibrium free energy at finite \( N \) takes the form
 
\begin{equation*}
	G(\sigma,\beta) = -\frac{1}{\beta }\log\left\{\sum_{p}\exp\left[-\beta N \,\bar{g}(p;\sigma,\beta)\right]\right\},
\end{equation*}
 see the thin lines in Fig.~\ref{fig:MarginalFreeEnergyFiniteN}(first row).
 The equilibrium tension-elongation relation is obtained by differentiating the free energy with respect to the loading \( \Sigma \), 
\begin{equation}
	\label{eq:average_y}
	\mean{y}(\sigma,\beta) = -\frac{\partial}{\partial \Sigma}G(\sigma,\beta) = \sigma-\mean{p}(\sigma,\beta).
\end{equation}
The resulting tension-elongation relations are shown  for different system sizes in Fig.~\ref{fig:MarginalFreeEnergyFiniteN}(second row, thin lines).

To address the thermodynamic limit we  introduce the equilibrium free energy per cross-linker, \( g=G/N \). When $N \to \infty$, we can use Laplace method to obtain
\begin{equation*}
	g(\sigma,\beta)_{\infty} =  \bar{g}_{\infty}\left[p_{*}(\sigma,\beta);\sigma,\beta\right],
\end{equation*}
where \( p_{*} \)  has been defined above as the minimum of \( \bar{g}_{\infty} \) at a given \( \sigma \).
In the same limit  we obtain 
$
	\mean{y}(\sigma,\beta) = y_{*}(\sigma,\beta),
$
where \( y_{*} = \sigma-p_{*} \)  is  the solution of
$ \sigma (y,\beta) - \sigma=0$
with
\begin{equation*}
	\label{eq:sigma_hard_device}
	\sigma(y,\beta) 	=y+\frac{1}{2}\left\{1-\tanh\left[\frac{\beta}{2}(y-y_0)\right]\right\}.
\end{equation*}
 
 The equilibrium tension-elongation relations are illustrated  in Fig.~\ref{fig:Average_configuration}(b). These curves are monotone when \( \beta\leq 4 \), while for \( \beta>4 \), they exhibit a plateau that extends as temperature decreases.
 In the limit \( \beta\to\infty \), we recover the equilibrium response of the purely mechanical system corresponding to the global minimum path.
 The three points A, B and C represent the three solutions of \( \sigma(y,\beta)=\sigma_0 \): the system switches collectively between the fully synchronized configurations A and C.
The transformation at the microscale associated with such switches translates into a large length change, whose amplitude is easily detectable even in the presence of thermally induced fluctuations. 
The device is then able to  deliver a strong mechanical output in response to small load perturbations. 

The behavior of the equilibrium \emph{stiffness} of a parallel bundle of switching elements  is discussed in \ref{sec:stiffness}.
In   \ref{sec:thermal_behavior} we study the \emph{thermal} properties of this system and  show that below the critical temperature, not only isothermal but also adiabatic response of this Brownian snap spring is  abrupt, and is accompanied by a substantial heat  release in response to small variations of the external force.
 

\section{Critical behavior} 
\label{sec:critical_behavior}

We have shown in the previous section that above the critical temperature our switching elements are completely desynchronized.
This leads to a smooth mechanical response while the snap spring property is lost. Instead,  below the critical temperature, the system can sustain thermal fluctuations and preserve its bistability.
The latter can be then viewed as a cooperative effect.

At the critical point \( \beta=4,\,\sigma=\sigma_0 \), the system has a degenerate  energy landscape with a marginally stable minimum corresponding to the configuration with \( p=1/2 \); see Fig.~\ref{fig:Average_configuration}[(a) dashed line]. Near the critical point    we obtain the  expansions
 $
 	p\sim1/2\pm (\sqrt{3}/4)[\beta-4]^{1/2},
 $
  for \( \sigma=\sigma_{0} \), and 
 $
 	p\sim1/2 - \mathrm{sign}[\sigma-\sigma_0]\left[(3/4)\left|\sigma-\sigma_0\right|\right]^{1/3},
$
for \( \beta=4 \).
Both  critical exponents  take the classical (Landau) values characteristic of the mean field Ising model \citep{Balian:2006wd}.
These results suggest that small deviations from the critical point lead  to significant changes in the configuration of the system.
In \ref{sec:stiffness}, we study  the behavior of thermal and mechanical susceptibilities and show  they reach their maximal values at \( \sigma=\sigma_0 \) and diverge at   the critical point, see Fig.~\ref{fig:susceptibility}.
 
  Near the critical point we can write
\(
	\mean{y}-y_0 = \pm (\sqrt{3}/4)\left[\beta-4\right]^{1/2}
\),
for \( \sigma=\sigma_{0} \),  and  
\(
	\mean{y}-y_0 =  \mathrm{sign}[\sigma-\sigma_0]\left[(3/4)\left|\sigma-\sigma_0\right|\right]^{1/3}
\), for \( \beta=4 \).
The latter relation suggests that in  critical conditions the   tension-elongation curve has a plateau; see dashed line in Fig.~\ref{fig:Average_configuration}(b).  The corresponding stiffness is equal to 0 , see \ref{sec:stiffness},  which means that upon application of a small positive (respectively negative) force increment, the system rapidly   unfolds (respectively folds).
Note that, below the critical temperature, similar behavior becomes irreversible, \emph{i.e} depending on the position on the hysteresis loop the system can respond abruptly to  either increase or a decrease of the external force.
Characteristically, near the critical point both large elongation and large shortening can take place, which makes our system very reactive and versatile.


\section{Rate dependent behavior} 
\label{sec:dynamic_loading}
\subsection{Single element} 
\label{sub:single_element}

\begin{figure}
	\centering
	\includegraphics{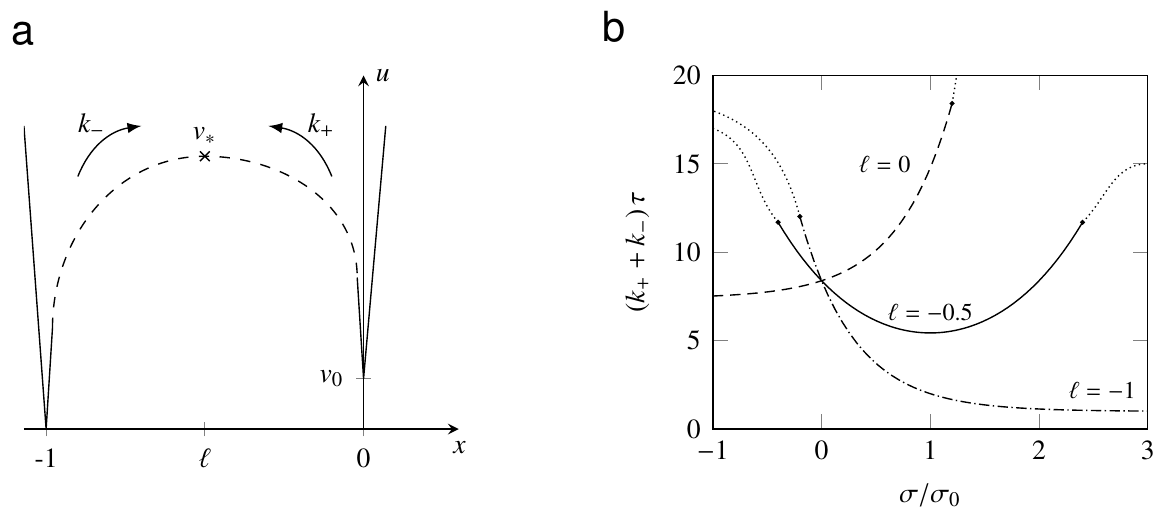}
	\caption{
	(a) Generalization of the Huxley and Simmons model of the energy barriers involving a transition state \( v_{*} \) corresponding to the conformation \( \ell \). (b) Equilibration rate between the states as function of the loading parameter at different values of \( \ell \). In (b) \( v_0=1 \), \( v_{*}=1.2 \) and \( \beta=2 \). Dotted constructions show a schematic representation of diffusion (\emph{vs.} reaction) dominated processes.
	}
	\label{fig:TransitionRates}
\end{figure}
In some regimes the isothermal  kinetics  of a single  switching element can be modeled using the Kramers approximation, which stipulates that the transition rate between two states depends primarily on  the   energy barrier separating these states. 
An expression of the energy barrier can be obtained  by generalizing the original rough HS model. To this end we define a transitional state \( x=\ell\in[-1,0]\), between the unfolded (\( x=0 \)) and the folded (\( x=-1 \)) configurations, characterized by the energy \( v_{*}>v_0 \)
, see Fig.~\ref{fig:TransitionRates} cf. \cite{Bormuth:2014hh}.
In a soft device the kinetic response of a single element under fixed load \( \sigma \) includes the discrete jumps associated with \( x \), and the continuous dynamics of \( y \).
We  assume for simplicity that the viscous relaxation timescale associated with the elongation described by the variable \( y \) is negligible, which implies that \( y \) can be adiabatically eliminated and replaced by its equilibrium value at a given  \( \sigma \), namely \( y=\bar{y}=\sigma+x \).
This assumption can be of course relaxed if we  endow the variable \( y \) with the appropriate Langevin dynamics.

In this simplified framework, we introduce the  chemical rate constants describing the load-dependent kinetics  of the direct and reverse transitions
\begin{equation*}
	\begin{split}
		\mathbb{P}(x^{t+dt}&=-1\,|\,x^{t}=0) = k_{+}(\sigma,\beta)\,dt,\\
		\mathbb{P}(x^{t+dt}&=0\,|\,x^{t}=-1) = k_{-}(\sigma,\beta)\,dt,
	\end{split}
\end{equation*}
where both \( k_{+} \) and  \( k_{-} \) depend on the loading \( \sigma \) and temperature $\beta$.
By taking into account the elastic element, the energy barriers for the forward and the reverse transition can be written as
\begin{equation*}
	\begin{split}
		\Delta u_{+}(\sigma,\beta) &= v_{*} - \ell \sigma - v_0, \\
		\Delta u_{-}(\sigma,\beta) &= v_{*}  - \ell\sigma -\sigma,
		\end{split}
\end{equation*}
and the corresponding transition rates are 
\begin{equation}
	\label{eq:transition_rates_function_of_sigma}
	\begin{split}
		k_{+}(\sigma,\beta) &= (1/\tau)\exp\left[\beta\,\left(\ell\,\sigma + \sigma_0\right)\right],\\
		k_{-}(\sigma,\beta) &= (1/\tau)\exp\left[\beta\,\left(\ell\, \sigma + \sigma \right)\right],
	\end{split}
\end{equation}
where \( 1/\tau = \alpha \exp[-\beta\,v_{*}]\), with \( \alpha=\text{const.} \), determines the timescale of the response.
Notice that the above expressions are valid if \( \Delta u_{+} \) and \( \Delta u_{-} \) are positive \emph{i.e.} for \( (v_{*}-v_0)/\ell < \sigma < v_{*}/(\ell+1)\).
Beyond this interval the Kramers description of the kinetics is no more valid and the dynamics response is controlled by the diffusive relaxation of the variables \( x \), so one can expect a saturation at large loads, see \cite{Bell:2017ct}.
In the rest of the section, we only consider the situation where Eq.~\eqref{eq:transition_rates_function_of_sigma} is valid.
In their original paper, HS worked in a particular setting where \( \ell=-1 \) which implies that \( k_{-}=\text{const.}\) and \( k_{+} (\sigma,\beta) = {k_{-}}\exp\left[-\beta\left(\sigma-\sigma_0\right)\right] \).

The equilibration rate \( k_+ + k_- \) is represented as a function of the normalized load \( \sigma/\sigma_0 \) in Fig.~\ref{fig:TransitionRates}(b). In the two limit cases \( \ell=0 \) or \( -1 \), this rate is monotone and converges to constant values in the limits \( y\to\pm\infty \).
In the case \( \ell=-1 \) considered by HS,  the rate is a decreasing function of the load, see Fig.~\ref{fig:TransitionRates}(b), dash-dotted line.
This conclusion has been challenged by recent experimental data showing a nonmonotone dependence of the rate on the load, which supports the present generalization of the HS model, see Fig.~\ref{fig:TransitionRates}(b), solid line and \cite{Offer:2016fn}. 
Moreover, the new experimental data show a saturation of the equilibration rate at large loads, which corroborates the fact that Kramers description is restricted to a finite interval of loads beyond which the dynamics is controlled by diffusion, see the dotted lines illustrating this saturation in Fig.~\ref{fig:TransitionRates}(b).

\subsection{Parallel bundle of \( N \) elements} 
\label{sub:parallel_bundle_of_n_elements}

\subsubsection{Kinetic model} 
\label{ssub:kinetic_model}

\begin{figure}
	\centering
	\includegraphics{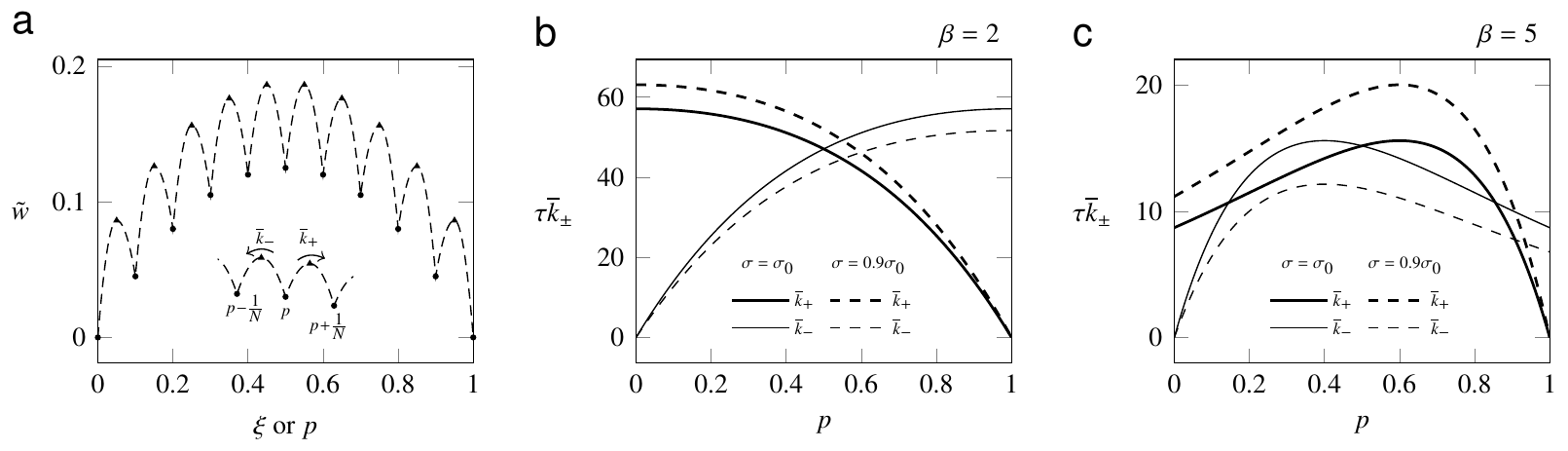}
	\caption{Kinetic model for a parallel bundle of \( N \) elements. (a) Energy landscape characterizing the sequential folding process of \( N=10 \) bistable elements at \( \sigma=\sigma_0 \). [(b) and (c)] Transition rates represented for \( \sigma=\sigma_0 \) (solid lines) and \( \sigma=0.9\sigma_0 \) (dashed lines) at \( \beta=2 \) (b) and \( \beta=5 \) (c). Parameters are \( v_0=1 \), \( v_{*}=1.2 \), and \( \ell=-0.5 \). In (a) \( \beta=2 \).
	}
	\label{fig:EnergyLandscapeWithBarriers}
\end{figure}

To characterize the transition rates in a cluster of \( N>1 \) elements under fixed external force, we introduce the energy \( \tilde{w}(p,p_{*}) \) corresponding to a configuration where \( p \) elements are folded (\( x_i=-1 \)) and \( p_{*} \) elements are at the transition state (\( x_i=\ell \)). The ``energy landscape'' separating two configurations \( p \) and \( q \) can then be expressed in terms of the ``reaction coordinate'' \( \xi = p-x(q-p) \), see Fig.~\ref{fig:EnergyLandscapeWithBarriers}(a).

The transition rates between neighboring metastable states are controlled by the energy barriers \( \Delta w_{+}(\sigma) = \tilde{w}(p,1/N;\sigma)-\bar{w}(p;\sigma)\) and \( \Delta w_{-} = \tilde{w}(p-1/N,1/N;\sigma)-\bar{w}(p;\sigma) \), where \( \bar{w} \) is defined by Eq.~\eqref{eq:hat_w}. We have
\begin{equation*}
	\begin{split}
		N\Delta w_{+}(p;\sigma) &= v_{*} - \ell (\sigma - p) - \sigma_0 + (1+3/N)(\ell^{2}/2)\\
		N\Delta w_{-}(p;\sigma) &= v_{*} - (\ell+1) (\sigma - p)  +(1+3/N)(\ell^{2}/2) -(1+N+2\ell)/(2N)
	\end{split}
\end{equation*}
which shows that the barriers corresponding to individual transitions are of the order of \( 1/N \) and therefore vanish in the thermodynamic limit; see a more detailed discussion of the energy barriers in \cite{Caruel:2015im}.

By taking into account the degeneracy of the metastable states, we obtain that the ensuing one-dimensional random walk process is characterized by  the transition probabilities 
\begin{equation}
	\label{eq:markov_random_walk}
	\begin{split}
 		\mathbb{P}\left[p^{t+dt} = p^{t}+1/N\right]  &= \bar{k}_{+}(p^{t};\sigma,\beta) \,dt,\\
		\mathbb{P}\left[p^{t+dt} = p^{t}-1/N\right] &= \bar{k}_{-}(p^{t};\sigma,\beta)\,dt,\\
		\mathbb{P}\left[p^{t+dt} = p^{t}\right] &= 1- 
		\mathbb{P}\left[p^{t+dt} = p^{t}+1/N\right] - 
		\mathbb{P}\left[p^{t+dt} = p^{t}-1/N\right],
	\end{split}
\end{equation}
where
\begin{equation}
	\label{eq:k_plus_k_minus}
	\begin{split}
		\tau\,\bar{k}_{+}(p;\sigma,\beta) &= 
		N\,(1-p)\,\exp\left\{-\beta \left[- \ell (\sigma - p) - \sigma_0 + (1+3/N)(\ell^{2}/2)\right]\right\},\\
		\tau\,\bar{k}_{-}(p;\sigma,\beta) &= N\,p\,\exp\left\{
		-\beta\left[- (\ell+1) (\sigma - p)  +(1+3/N)(\ell^{2}/2) -(1+N+2\ell)/(2N)\right]\right\}.
	\end{split}
\end{equation}
Behind the construction of this Markov chain is the assumption that only one element can switch within the interval \( [t,t+dt] \).
This assumption is reasonable as one can show that the energy barriers corresponding to several switching elements (\( p_{*}>1/N \)) are larger than the barrier involving only one shifting element.

The nondimensional rates \eqref{eq:k_plus_k_minus} are represented in Fig.~\ref{fig:EnergyLandscapeWithBarriers}[(b) and (c)] for two different loading conditions, and two different temperatures. 
Notice that in Eq.~\eqref{eq:k_plus_k_minus}, the prefactor and the exponential term have opposite tendencies in terms of \( p \).
At large temperatures (\( \beta\to 0 \)) the entropic prefactor dominates and the rates are monotone functions of \( p \), see Fig.~\ref{fig:EnergyLandscapeWithBarriers}(b).
At low temperatures, the exponential terms become non-negligible and the rate dependence on \( p \) looses monotonicity, see Fig.~\ref{fig:EnergyLandscapeWithBarriers}(c).

The mechanical coupling appearing in the exponent of \eqref{eq:k_plus_k_minus} makes the dynamics non-linear.
We recall that in the original HS model the dependence of the transition rates on \( p \) was linear because the cross-bridges were viewed as completely independent.
 The master equation describing the evolution of the probability density  (omitting the dependence on \( \sigma \) and \( \beta \) for clarity) reads,
\begin{equation}
	\label{eq:master_equation}
	\frac{\partial}{\partial t}\bar{\rho}(p,t) = \bar{k}_{+}\left(p-1/N\right) \bar{\rho}\left(p-1/N,t\right)
	+ \bar{k}_{-}\left(p+1/N\right)
	\bar{\rho}\left(p+1/N,t\right)
	- \left[\bar{k}_{+}(p)+\bar{k}_{-}(p)\right]\bar{\rho}\left(p,t\right).
\end{equation}

The  time independent equilibrium  distribution, which can be obtained explicitly as a  steady state solution \( \bar{\rho}_{eq} \) of Eq.~\eqref{eq:master_equation}, takes the form \citep{vanKampen:2001vs}
\begin{equation*}
	\bar{\rho}_{eq}(p) = \frac{
	\prod_{Ni=1}^{Np}\frac{\bar{k}_{+}(i-1/N)}{\bar{k}_{-}(i)}
	}{
	1+\sum_{Nj=0}^{N}\prod_{Ni=0}^{Nj}\frac{\bar{k}_{+}(i-1/N)}{\bar{k}_{-}(i)}
	}.
\end{equation*}
Using the definitions \eqref{eq:k_plus_k_minus}, we obtain after simplifications
\(
	\bar{\rho}_{eq}(p;\sigma,\beta) = Z(\sigma,\beta)^{-1}\exp\left[-\beta N \bar{g}(p;\sigma,\beta)\right],
\)
which is in full agreement with Eq.~\eqref{eq:rho_bar}.


\subsubsection{First passage times and equilibration rates} 
\label{ssub:first_passage_times_and_equilibration_rates}

To understand the peculiarities of the time dependent response  of the parallel bundle of $N$ elements, it is instructive to  first derive the expression for the mean first passage time \( \tau(p\to p') \)  characterizing transition between two metastable states with \( Np \) and \( Np' \) (\( p<p' \), the analysis in the case \( p>p' \) is similar) switched elements. Here assume a reflective boundary condition at \( p=0 \) and an absorbing boundary condition at \( p' \). Following \cite{Gardiner_2004} and omitting again the dependence on \( \sigma \) and \( \beta \), we can write
\begin{equation}
	\label{eq:first_passage_time}
	\tau(p\to p') = \sum_{Nk=Np}^{Np'}\left[\bar{\rho}\left(k\right)\,
		\bar{k}_{+}\left(k\right)\right]^{-1}\sum_{Ni=0}^{Nk}\bar{\rho}\left(i\right),
\end{equation}
where \( \bar{\rho} \) is the marginal equilibrium distribution at fixed \( p \) given by Eq.~\eqref{eq:rho_bar}, and \( \bar{k}_{+} \) is the forward rate \eqref{eq:k_plus_k_minus}.

\begin{figure}
	\centering
	\includegraphics{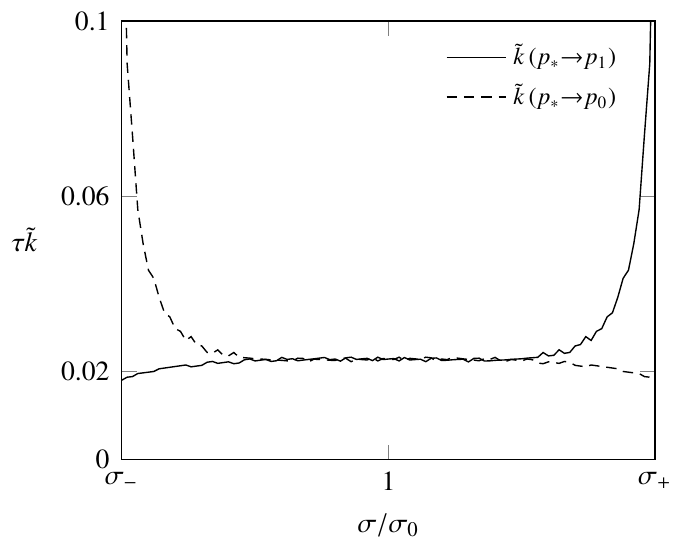}
	\caption{
	Characteristic rate of relaxation to the metastable state in the case of a reflecting barrier at \( p=\hat{p} \). The parameters are \( N=200 \), \( \beta=5 \) and \( \ell=-0.5 \).
	}
	\label{fig:InnerBassinRate}
\end{figure}

In the case \( \beta>\beta_c \)
for the interval of loading \( [\sigma_{-},\sigma_{+}] \), the marginal free energy \( \bar{g} \) has two minima which we denote \( p=p_{1} \) and \( p=p_{0} \),  with \( p_{0}<p_{1} \). The minima are  separated by a maximum located at \( p=\hat{p} \). 

 We first illustrate the relaxation rate inside one of the macroscopic wells by establishing a reflective boundary condition at \( p=\hat{p} \).
The average rates \( \tilde{k}(\hat{p}\to p_{1,0}) = 1/\tau(\hat{p}\to p_{0,1}) \) of reaching the metastable states (\( p=0 \) or \( p=1 \)) starting from the top of the energy barrier \( \hat{p} \) are represented as functions of the load in Fig.~\ref{fig:InnerBassinRate}.
These rates are almost constant over the entire interval of existence of the metastable states \( [\sigma_-,\sigma_+] \).
The rise at the boundaries is due to the proximity of the barrier to the bottom of the energy well.
From Fig.~\ref{fig:InnerBassinRate} we conclude, for instance, that for \( N=200 \) and \( \beta=5 \), the timescale of the internal relaxation is of the order of \( 10^{-2} \).

If we now consider the transition between the two macrostates, Eq.~\eqref{eq:first_passage_time} can be simplified
if \( N \) is sufficiently large.
In this case, the sums in Eq.~\eqref{eq:first_passage_time} can be transformed into integrals  
 \begin{equation}
	 \label{eq:first_passage_time_integrals}
 	\tau(p_0\to p_{1}) = N^{2}\int_{u=p_0}^{p_{1}}\left[\bar{\rho}(u)\bar{k}_{+}(u)\right]^{-1}\left[\int_{0}^{u}\bar{\rho}(v)dv\right]du.
 \end{equation}

The inner integral in Eq.~\eqref{eq:first_passage_time_integrals} can be computed using Laplace method and noticing that the function \( \bar{g} \) has a single minimum in the interval \( [0,u>p_0] \) located at \( p_{0} \), we have
\begin{equation*}
	\tau(p_{0}\to p_{1}) = \sqrt{2\pi N} \left[\beta\,\left|\bar{g}''(p_0)\right|\right]^{-\frac{1}{2}}\int_{u=p_0}^{p_{1}}\left[\bar{\rho}(u)\,\bar{k}_{+}(u)\right]^{-1}\bar{\rho}\left(p_{0}\right)du.
\end{equation*}
In the remaining integral, the inverse density \( \left(1/\bar{\rho}\right) \) is sharply peaked at \( p=\hat{p} \) so again using Laplace method we can write
\begin{equation}
	\label{eq:T_non_convex}
	\tau(p_{0}\to p_{1}) =
	2\pi\left(N/\beta\right)\bar{k}_{+}(\hat{p})^{-1}\left|\,\bar{g}''(p_0)\,\bar{g}''(\hat{p})\,\right|^{-\frac{1}{2}}\exp\left\{\beta\,N\left[\bar{g}^{\infty}(\hat{p})-\bar{g}^{\infty}(p_0)\right]\right\},
\end{equation}

\begin{figure}
	\centering
	\includegraphics{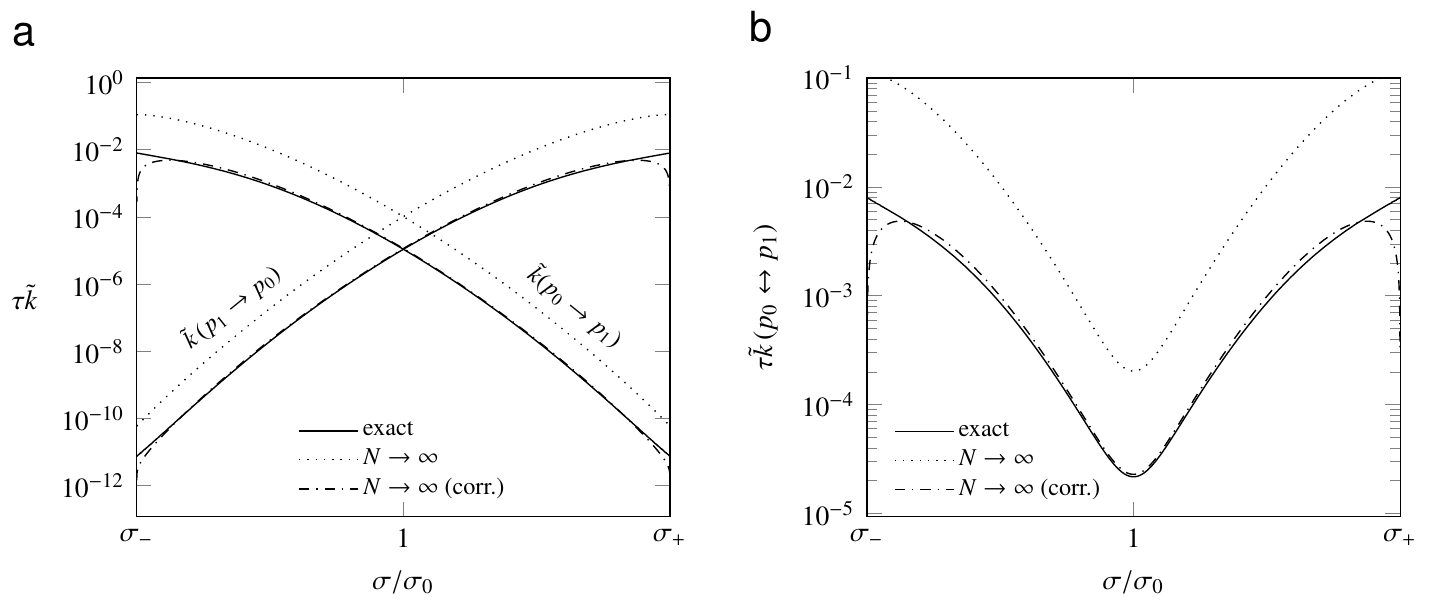}
	\caption{
	Transition rates between the two macroscopic configurations \( p_0(\sigma) \) and \( p_1(\sigma) \). (a) Forward [\( \tilde{k}(p_0\to p_1) \)] and reverse [\( \tilde{k}(p_1\to p_0) \)] rates. (b) Equilibration rate \( \tilde{k}(p_0\leftrightarrow p_1) = \tilde{k}(p_0\to p_1) + \tilde{k}(p_1\to p_0) \). Solid line, exact computation from Eq.~\eqref{eq:first_passage_time}; dotted line, thermodynamic limit approximation given by Eq.~\eqref{eq:T_non_convex}; dot-dashed line, thermodynamic limit with corrected free energy. Parameters are the same as in Fig.~\ref{fig:InnerBassinRate}.
	}
	\label{fig:transitionTimeScale}
\end{figure}

The rates \( \tilde{k}(p_{0,1}\to p_{1,0}) = \left[\tau\left(p_{0,1}\to p_{1,0}\right)\right]^{-1} \) corresponding to the forward (\( p_0\to p_1 \)) and reverse (\( p_1\to p_0 \)) collective transitions are represented in Fig.~\ref{fig:transitionTimeScale}(a) as function of the loading \( \sigma\in[\sigma_{-},\sigma_+] \) for \( N=200 \).
The presence of the barrier considerably slowers the transition process comparing to the simple relaxation process illustrated in Fig.~\ref{fig:InnerBassinRate}. However, the order of magnitude (\( 10^{-2} \)) of the simple relaxation process is recovered at the boundaries of the interval \( [\sigma_-,\sigma_+] \).

The comparison of the intrabassin relaxation (see Fig.~\ref{fig:InnerBassinRate}) and the interbassin transition (see Fig.~\ref{fig:transitionTimeScale}) shows that the system has now two characteristic timescales. The first one depends on the micro energy barriers between neighboring micro-configurations and scales with \( \tau \).
 The second timescale is of the order of \( \exp[ N \Delta \bar{g}] \) [see Eq.~\eqref{eq:T_non_convex}], where \( \Delta\bar{g} \) is the height of the macroscopic energy barrier separating the two metastable states.
  In the thermodynamic limit, this energy barrier grows linearly with \( N \), which prevents collective \emph{interbasin} dynamics  and generates metastability: glass like  ``freezing'' in one of the two globally coherent configurations.

Note that in Fig.~\ref{fig:transitionTimeScale} there is a non-negligible quantitative difference between the  exact computation based on Eq.~\eqref{eq:first_passage_time}(solid line) and the asymptotic formulas derived in Eq.~\eqref{eq:T_non_convex} (dotted line).
This difference may be explained by the fact that even though the microscopic barriers vanish in the thermodynamic limit, the dynamics cannot be straightforwardly reduced to a diffusion process on the smooth potential \( \bar{g}_{\infty} \). The presence of at least two scales in the kinetics can lead to more subtle asymptotic description, see for instance \cite{Zwanzig:1988vl}.

One can refine the analytical results by considering a finer approximation of the entropy in the thermodynamic limit. In Eq.~\eqref{eq:g_bar_y_infinity}, the ideal mixing entropy \( s_{\infty} = -p\log(p) - (1-p)\log(1-p) \) is obtained from the Stirling formula \( \log(n!) = n\log(n)-n + O[\log(n)]\). A finer approximation consists in writing \( \log(n!) = n\log(n)-n + (1/2)\log(2\pi n) + O(1/n)\), which leads to \( s_{\infty} = -(p+\frac{1}{2N})\log(p) - (1-p+\frac{1}{2N})\log(1-p) - \frac{1}{2N}\log(2\pi N) \).
By Using the latter expression for the entropy in the definition of the marginal free energy \eqref{eq:g_p_infinity} we can obtain a better quantitative agreement between the exact and the approximated rates, see dot-dashed lines in Fig.~\ref{fig:transitionTimeScale}.
In conclusion we observe that the above computations suggest that the equilibration rate between the two macrostates [see Fig.~\ref{fig:transitionTimeScale}(b)] decreases exponentially in the vicinity of \( \sigma=\sigma_0 \) while increasing for both \( \sigma>\sigma_0 \) and \( \sigma<\sigma_0 \).
This contradicts the non symmetric behavior of the transition rate predicted in \cite{Huxley_1971} where the symmetry was broken due to the assumption \( \ell=-1 \). Our more general results for \( -1\leq\ell\leq 0 \) are supported by the recent experimental evidence reported in \cite{Offer:2016fn}.


\subsubsection{Coarse grained dynamics} 
\label{ssub:coarse_grained_dynamics}

The simplest approach to solving Eq. \ref{eq:master_equation}  is focusing exclusively on the dynamics of averages. To this end we may multiply Eq.~\eqref{eq:master_equation} by \( p \) and sum the result over \( p \).  We then obtain a   kinetic equation
\begin{equation}
	\label{eq:kinetic_equation}
	N\frac{\partial}{\partial t} \mean{p} = 
	\mean{\bar{k}_{+}\left(p\right)} - \mean{\bar{k}_{-}\left(p\right)}.
\end{equation}
To close this model we need to use the  mean-field type approximation giving us the (nonlinear) analog of the chemical reaction equation of HS; see \cite{Huxley_1971}. By noticing that the rate functions are convex in \( p \)  we can expand them around \( \mean{p} \) to obtain,
\[
\mean{\bar{k}_{\pm}(p)} = \bar{k}_{\pm}(\mean{p}) + \frac{1}{2}\mean{\left[p-\mean{p}\right]^{2}}\left.\frac{\partial^{2} \bar{k}_{\pm}}{\partial p^{2}}\right|_{\mean{p}} + o\left\{\mean{\left[p-\mean{p}\right]^{2}}\right\}.
\]
In the thermodynamic limit, the fluctuations can be neglected and the kinetic equation \eqref{eq:kinetic_equation} takes the mean-field form
\begin{equation}
	\label{eq:mean_field_dynamics}
	N\frac{\partial}{\partial t} \mean{p} = \bar{k}_{+}\left(\mean{p}\right) - \bar{k}_{-}\left(\mean{p}\right).
\end{equation}
In \ref{sec:stiffness}, we show that   close to the critical point we have the divergence \( \mean{\left[p-\mean{p}\right]^{2}}\sim\left[\beta - 4\right]^{-1} \), which limits the applicability of the approximate equation \eqref{eq:mean_field_dynamics}.

To follow  the evolution of the variable $\mean{y}(t)$ in response to
 a prescribed loading protocol  \( \sigma (t)\), we can use the equilibrium relation \( \mean{y}=\sigma-\mean{p} \) to obtain  
$
	\dot{\mean{y}} = \dot{\sigma} + (1/N)\left(\bar{k}_{-}(\sigma-\mean{y})- \bar{k}_{+}(\sigma-\mean{y}\right).
$ The steady state solution of \eqref{eq:mean_field_dynamics} can be  then computed as a root of the transcendental equation
\begin{equation*}
	\mean{p}_{\infty} =  \frac{1}{2}\left\{1-\tanh\left[\frac{\beta}{2}\left(\sigma-\sigma_{0}-\mean{p}_{\infty}+\frac{1}{2}\right)\right]\right\}.
\end{equation*}



\begin{figure*}[t]
	\centering
	\includegraphics[]{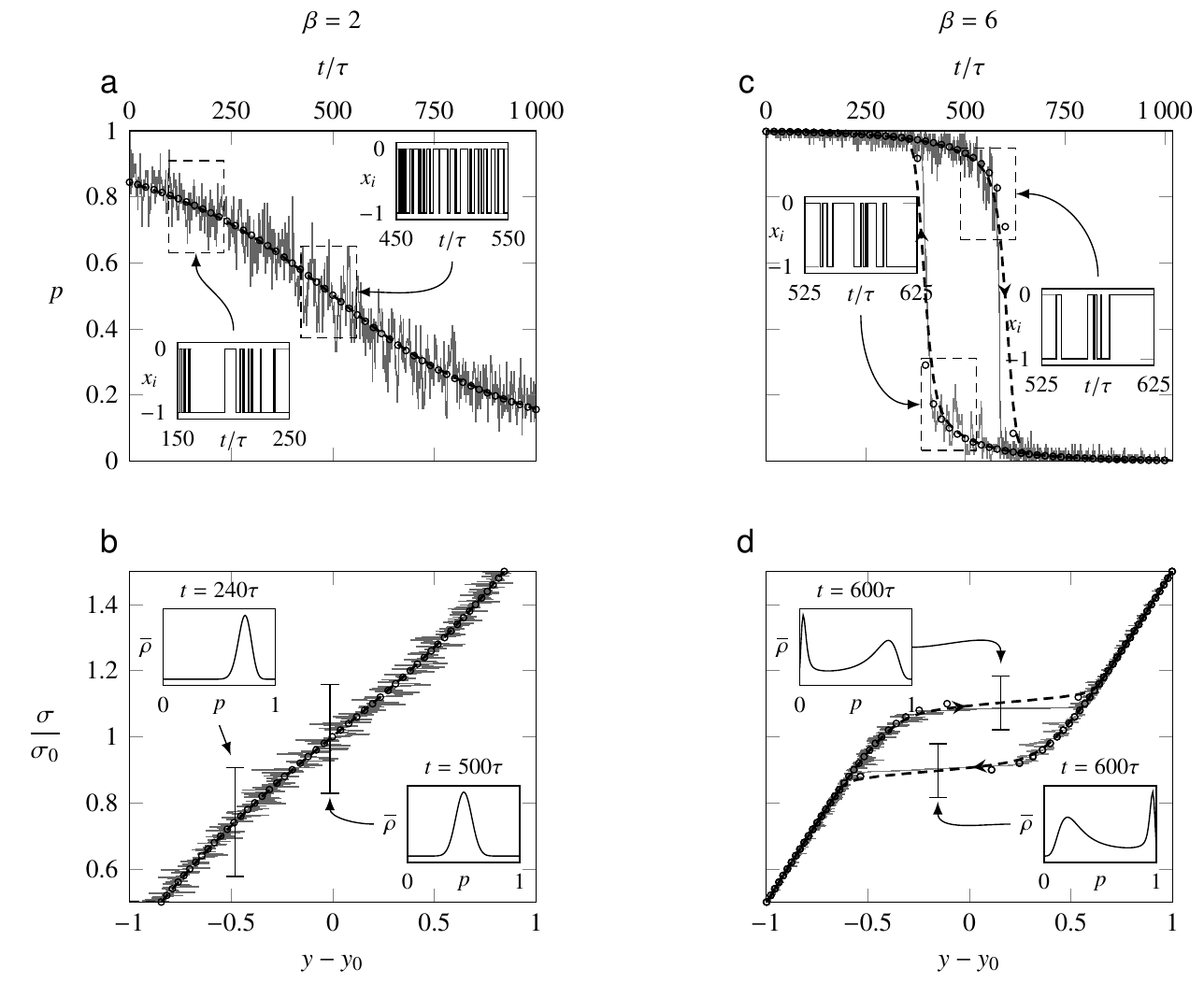}
	\caption{
	Stochastic response to a quasi-static tension ramp imposed on a system with \( N=100 \) and \( \beta=2 \) (a,b) or \( \beta=6 \) (c,d). In (a) and (b), the loading consists in a tension ramp from \( 0.5\sigma_0 \) to \( 1.5\sigma_0 \) achieved within \( 10^3\tau \). In (c) and (d), two trajectories are shown corresponding to two ramps, one from \( 0.5\sigma_0 \) to \( 1.5\sigma_0 \) and the other from \( 1.5\sigma_0 \) to \( 0.5\sigma_0 \). (a) and (c) Stochastic evolution [obtained from Eq.~\eqref{eq:markov_random_walk}] of the fraction of folded snap springs (gray) together with averages computed from the simulation of the master equation \eqref{eq:master_equation} (dashed lines). 
	(b) and (d) Trajectories (gray) and corresponding averages (dashed lines) in the \( (y,\sigma) \) space.
	The inserts show samples of the trajectory of a single element corresponding to different time windows [(a) and (c)] and  snapshots of the probability density \( \bar{\rho} \) obtained from the numerical simulation of \eqref{eq:master_equation} [(b) and (d)].
	Open symbols, responses obtained using the mean-field kinetic equation \eqref{eq:mean_field_dynamics}.
	}
	\label{fig:quasi_static_trajectories}
\end{figure*}

\subsubsection{Numerical illustrations} 
\label{ssub:numerical_illustrations}

As a first illustration of the obtained relations, we consider   behavior of the system with \( N=100 \) elements subjected to slow  increase  of force  from the value \( 0.5\,\sigma_0 \) to the value \( 1.5\,\sigma_0 \) in the time interval \( [0,1000\,\tau] \); see Fig.~\ref{fig:quasi_static_trajectories}. The trajectories were obtained using a simple first order Euler numerical scheme simulation of the Markov process \eqref{eq:markov_random_walk}, with a time step equal to \( 10^{-3}\tau \). Average trajectories and distributions were obtained by numerically solving Eq.~\eqref{eq:master_equation} with an explicit algorithm, which involved the computation of the transition matrix at each time step.

At supercritical temperatures [see (a) and (b)], the transition between the two states is smooth and can be described as a drift of  a  unimodal probability density, which means  that the system is fluctuating around a single mixed configuration with a time dependent  average.


Below the critical point, the response shows hysteresis; see Fig.~\ref{fig:quasi_static_trajectories} [(c) and (d)]. This phenomenon  is due to the presence of the macroscopic wells in the marginal free energy \( \bar{g} \).   The distribution of the elements inside the hysteresis  domain \( [\sigma_{-},\sigma_{+}] \) becomes bi-modal and  the escape time [see Eq.~\eqref{eq:T_non_convex}], 
characterizing the equilibration between the two wells grows as \( \exp[N] \).   This prevent the  equilibrium transition expected at \( \sigma=\sigma_0 \) to take place at the time scale of the loading. The collective switch is observed only when the energy barrier becomes of the order of \( 1 \), \emph{i.e.} when the loading parameter reaches the boundary of the bi-stable interval \( [\sigma_{-},\sigma_{+}] \).

We note  that the average trajectory is  captured rather well by both,  the   master equation (\ref{eq:master_equation}) and the mean-field kinetic equation (\ref{eq:mean_field_dynamics}),  even in the presence of the hysteresis loop. This is due to the fact that the loading is sufficiently slow and the system remains close to  the quasi-stationary states describing coherent  metastable configurations.


\begin{figure}[t]
	\centering
	\includegraphics[]{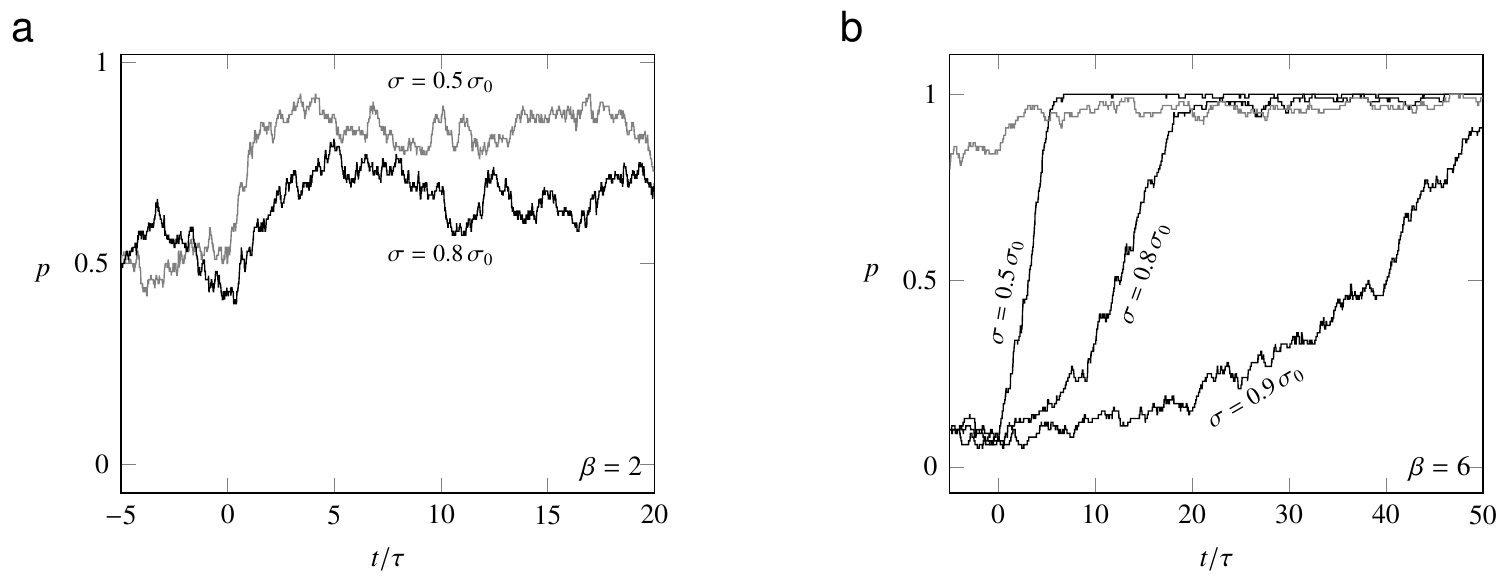}
	\caption{
	Trajectories obtained when submitting the system to an instantaneous tension drop applied at \( t=0 \) with \( \beta=2 \) (a) or \( \beta=6 \) (b).
The initial tension is set to \( \sigma=\sigma_0 \) and the different curves correspond to different tension step sizes. In (b) the initial state of the system is set at \( p=0.1 \) which corresponds to one of the macroscopic wells. The gray trace in (b) shows a single trajectory starting from the second macroscopic well (\( p=0.8 \)) for the case \( \sigma=0.9\sigma_0 \).
	}
	\label{fig:quick_recovery_trajectories}
\end{figure}


As a second illustration of the general theory, we now consider the kinetics of the relaxation of the system submitted to an instantaneous force drop. This process would mimic to the so-called phase 2 of the fast transient shortening observed in skeletal muscle fibers under force-clamp loading protocols \citep{Reconditi_2004}. 

We first show in Fig.~\ref{fig:quick_recovery_trajectories}  sample trajectories obtained for different values of the force drop at two different temperatures, \( \beta=2 \) [see (a)], and \( \beta=6 \); see (b). For \( \beta=2 \), the system is initiated in the configuration \( p=0.5 \), and the load is dropped  at \( t=0 \)  from \( \sigma_0 \) to \( 0.5\sigma_0 \) (low value) or \( 0.8\sigma_0 \) (high value).
For \( \beta=6 \) [see Fig.~\ref{fig:quick_recovery_trajectories}(b)], the system is initiated either at \( p=0.8 \) (gray trace) or at \( p=0.1 \) (black traces) and the load is dropped at \( t=0 \) from \( \sigma=\sigma_0 \) to \( 0.5\sigma_0 \), \( 0.8\sigma_0 \) or \( 0.9\sigma_0 \). 

First of all, for both  temperatures tested, we see that the larger the step size, the faster the response. 
At \( \beta=6 \), there are two metastable states for \( \sigma=\sigma_0 \), the first one with \( p\approx0.1 \) and the second one with \( p\approx 0.9 \). When the load is changed from \( \sigma=\sigma_0 \) to \( \sigma=0.9\,\sigma_0 \), there is only one remaining state with \( p\approx 1 \).
If the system is prepared in the well with local minimum at \( p=0.9 \) [see gray trace in Fig.~\ref{fig:quick_recovery_trajectories}(b) obtained with the initial condition \( p=0.8 \)], it relaxes to the final state \( p\sim 1 \), without crossing any macroscopic barrier and in this respect the response is analog to the case where the system is not bistable \emph{i.e.} it is characterized by only one timescale; see Fig.~\ref{fig:quick_recovery_trajectories}(a).
However, if the system is prepared in the metastable with \( p=0.1 \), the response contains two phases; see the trace with \( \sigma=0.9\sigma_0 \) in Fig.~\ref{fig:quick_recovery_trajectories}(b).
The first phase of the relaxation is slower than the second phase, and corresponds to the escape of the system from its initial metastable state.
 The second phase corresponds to the relaxation inside the final coherent state, which occurs without crossing any macroscopic barrier.

 \begin{figure}
 	\centering 	\includegraphics[]{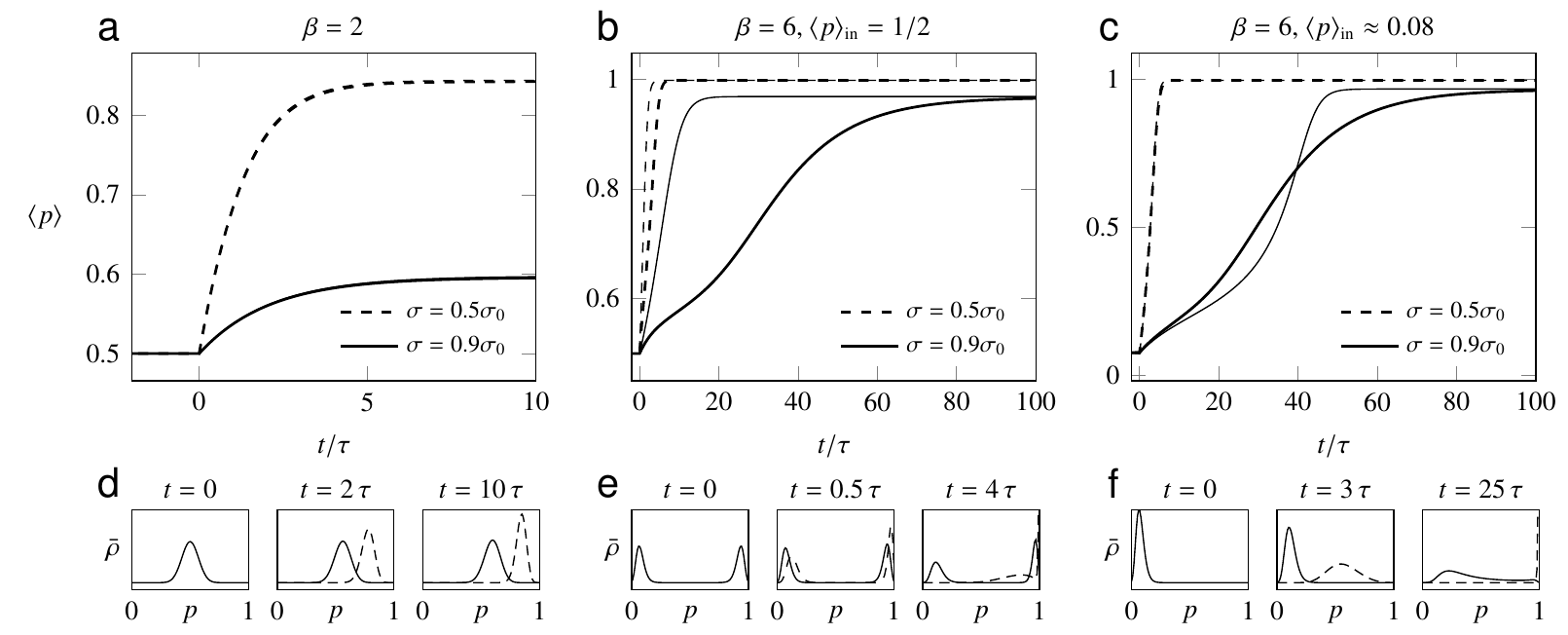}
 	\caption{
 	Relaxation of the average conformation in response to fast force drops applied at \( t=0 \) with \( \beta=2 \) (a) and \( \beta=6 \) [(b) and (c)]. Imposed forces after the step are equal to \( 0.5\sigma_0 \) (dashed), and \( 0.9\sigma_0 \) (solid). Thick lines, solutions from the master equation \eqref{eq:master_equation}; thin lines, solutions from the mean-field dynamic equation \eqref{eq:mean_field_dynamics}. 
 	In (b), the initial condition corresponds to thermal equilibrium: bimodal distribution and \( \mean{p}_{\text{\tiny in}}=1/2 \). In (c), the initial condition corresponds to the unfolded metastable state: unimodal distribution and \( \mean{p}_{\text{\tiny in}}\approx0.08 \).
 	Snapshots at different times of the probability density profiles corresponding to [(a)--(c)] are shown in [(d)--(f)], respectively.
 	}
 	\label{fig:quick_recovery_average_and_density}
 \end{figure}
 
To illustrate further the quick relaxation process we now  compare the average response obtained in the same loading conditions by using either   the master equation \eqref{eq:master_equation}, or the mean-field kinetic approximation \eqref{eq:mean_field_dynamics}; see Fig.~\ref{fig:quick_recovery_average_and_density}. We observe that the kinetic slowing down induced by the collective effects in the case of  high load, remains visible at low temperature: the corresponding probability distributions at different times are illustrated in  Fig.~\ref{fig:quick_recovery_average_and_density}(d--f).  Remarkably, the mean-field approximation is accurate at large temperatures [see thin lines in Fig.~\ref{fig:quick_recovery_average_and_density}(a)] while it fails to reproduce the exact dynamics at low temperatures, even though the final equilibrium states  are similar. 

The observed  difference between the chemo-mechanical description and the  simulations targeting the full probability distribution is due to the fact that for the mean-field kinetics the transition rates are  computed based on the average values of the order parameter. At large temperatures, where the distribution is uni-modal, the average value indeed correspond to the most  probable state and therefore the mean-field kinetic theory captures rather-well the timescale of the response; see Fig.~\ref{fig:quick_recovery_average_and_density} (a). Instead, at low temperatures, since the equilibrium distribution is bi-modal, the   average value     corresponds to a state which is very poorly  populated. The value of  the order parameter that actually makes the kinetics slow  corresponds   to a  metastable states and  because of this discrepancy the mean-field kinetic equation fails to reproduce the real dynamics of the system; see Fig.~\ref{fig:quick_recovery_average_and_density}[(b) and (e)].

Our results should not be interpreted in the sense that  the mean-field   approximation  \eqref{eq:mean_field_dynamics} fails to capture the two-stage relaxation process. We have just shown that at low temperatures   it takes a long time for the system  to  reach the  bimodal equilibrium distribution and   to acquire  the associated equilibrium average.  If instead the system is initially prepared  in one of the metastable configurations but with the average order parameter different from the corresponding equilibrium value,  the mean-field kinetics will capture  correctly  the two timescales of the response; see Fig.~\ref{fig:quick_recovery_average_and_density}(c).

\begin{figure}
	\centering
	\includegraphics{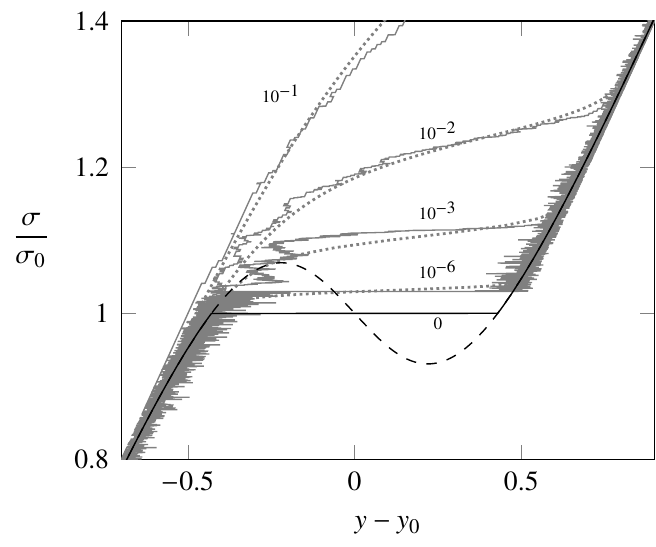}
	\caption{
	Response of the system to ramp loading at different rates \( \sigma_0/\tau\) (indicated by numbers besides each trace). Gray lines, stochastic trajectories (solid) shown together with average obtained from the master equation (dotted); black lines, metastable states (dashed) and equilibrium tension-elongation relation (thick solid). Parameters are \( N=100 \) and \( \beta=6 \).
	}
	\label{fig:RampAtDifferentRates}
\end{figure}

To further illustrate the kinetics of the system, we study next the dependence of the response on the rate of loading, see Fig.~\ref{fig:RampAtDifferentRates}.
 We vary the tension protocols in a broad range by changing the loading rate from \( 10^{-6}\,\sigma_0/\tau \) to \( 10^{-1}\,\sigma_0/\tau \).
In the limit of slow driving \( \sigma_0/\tau_0 \to 0\), the system can explore the whole phase space and therefore exhibits the equilibrium response, see solid black line in Fig.~\ref{fig:RampAtDifferentRates}.
At  faster loading rates (\( \sigma_0/\tau =10^{-3}\) of \( 10^{-6} \)) the metastability becomes an issue and the system develops hysteresis bounded by the spinodal states. In these   regimes the system remains trapped in the metastable states (dashed line) and the dynamics is dominated by interbasin kinetics, while only rare jumps between the metastable configurations are observed.
This behavior is exaggerated  as we  decrease the  temperature.
Finally,  sufficiently fast loading 
(\( \sigma_0/\tau =10^{-1}\) or \( 10^{-2} \)) drives the system away from the energy wells and 
the slowing down due to metastability becomes unimportant. In these regimes, the  energy landscape changes faster than the individual unfolding events and  the hysteresis becomes largely  controlled by the intrabasin relaxation time. 

Interestingly, quite similar behavior is exhibited by a chain of bistable elements connected in series, see \cite{Achenbach:1986kh,Efendiev_2010,Benichou:2015je} and \cite{Benichou:2016br}. A recent work suggests that the kinetics of successive switching events in such systems may be controlled by a single parameter linking the rate of loading and the rate of internal relaxation  \citep{Benichou:2015je,Benichou:2016br}. Similar data collapse can be expected to hold in the case of a parallel bundle of the HS type.

Our study of the kinetic properties of the model shows that the system has a mechanical memory due to the long escape time from a metastable state. 
When perturbed, it exhibits a relaxation rate that depends on the amplitude of the loading. The switch-like behavior characterized by a large mechanical output in response to small variations in external stress is then associated with a slow recovery phase, whose presence generates an effective damping which contributes to the robustness of the response in the presence of perturbations.


\section{Conclusions} 
\label{sec:conclusions}

The fact that a single mechanical snap spring loses its discontinuous response in the presence of  random perturbations  becomes an important issue   for the engineering devices employing bi-stability at nm scales. Behind the   problem of  building a ``brownian snap spring''   is the general problem of maintaining multi-stability in a fluctuating environment.
 An almost naive solution of this problem, proposed in the present paper,  is to use   a parallel cluster of \( N \) snap springs.
Despite its simplicity, this  mechanical toy,  exhibits a rich class of equilibrium mechanical behaviors including abrupt transitions  and criticality. 
 Our study of kinetics  shows that  even though the rate effects may interfere with the ability of the system  to undergo abrupt changes, such kinetic smoothening  takes place at the time scale of the transition itself and  if the controls are varied at the time scale which is larger than the fast  transition time, the temperature itself cannot prevent such device from undergoing quasi-discontinuous transitions that are now embedded into a hysteresis loop.
 
Our analysis helps to understand  why   the presence of long-range interactions in this system is crucial for the preservation of  bi-stability and ensuring  razor-edge response  at finite temperatures.   We show that the   ultra-sensitivity is achieved due to synchronization of  strongly interacting individual snap springs.    
The dependence of the critical temperature on the number of elements \( N \) suggests that the transition from  synchronized to de-synchronized behavior can be controlled by the system size.
At   the critical point, the mechanical response of the system  is characterized by a zero stiffness which makes it  anomalously responsive to external perturbations. This property  is highly desirable in applications because it    allows the system to amplify interactions and  ensure strong feedback. 

The prototypical  device  studied in the paper  can  serve as a mechanical memory unit  resistant to random  perturbations. The underlying macroscopic collective transition  has a strong thermal signature which reveals itself in a  substantial heat  release in response to the variation of the external force. This property could be useful for energy harvesting applications at the nanoscale. Another collective effect is the critical  slowing down near the critical point which can be used in the design of adaptive dampers.  The plethora of   other  applications for  systems  with bistable elements connected in parallel is discussed in \cite{Wu:2014dz,Wu:2015hi}. 
 
\section{Acknowledgments}

The authors thank P. Recho, R. Sheshka and F. Staniscia for helpful discussions and anonymous reviewers for constructive suggestions. LT was supported by the PSL grant BIOCRIT.

\appendix

\section{Susceptibility}
\label{sec:stiffness} 

In this Appendix we discuss the parametric dependence of the equilibrium susceptibility and stiffness in the case of a single switching element, and in the case of a parallel bundle of $N$ such elements.

We begin with the behavior of a single element discussed in Section \ref{sec:single_element}. The sensitivity of the system to changes in the external loading is characterized by the ``susceptibility''
\begin{equation}
	\label{eq:SingleSwitchChi}
	\chi(\sigma,\beta) = \frac{\partial}{\partial\sigma}\mean{x}(\sigma,\beta) = \beta \mean{\left[x-\mean{x}(\sigma,\beta)\right]^{2}},
\end{equation}
where 
\(
	\mean{\left[x-\mean{x}(\sigma,\beta)\right]^{2}} 
	=\left(1/4\right)\sech\left[\left(\beta/2\right)\left(\sigma-\sigma_0\right)\right]^{2},
\)
is the variance of $x$. 
\begin{figure}[]
	\centering
	\includegraphics[]{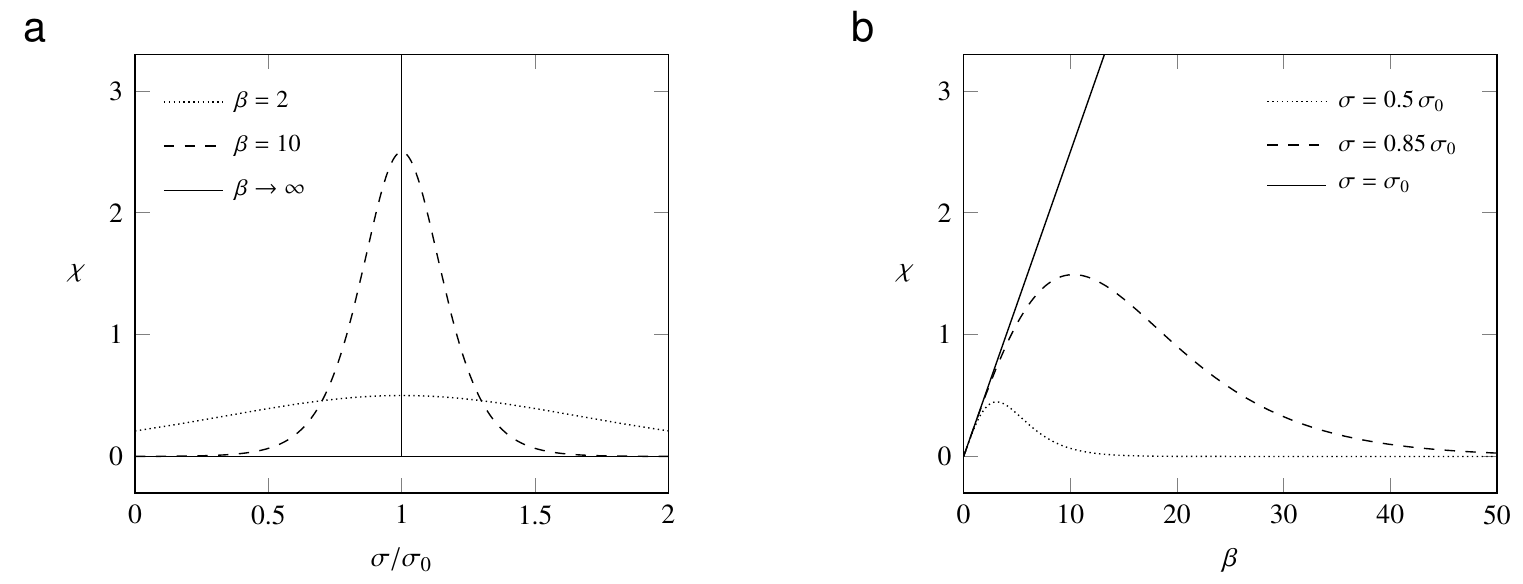}
	\caption{
	Susceptibility of a single digital switch represented as function of the loading (a), and as function of temperature (b).
	}
	\label{fig:SingleSwitchSusceptibility}
\end{figure}

As in the case of  a hard device [see \cite{Caruel:2016kw}], the susceptibility is maximum for \( \sigma=\sigma_0 \) [see Fig.~\ref{fig:SingleSwitchSusceptibility}(a)], and becomes infinite in the zero temperature limit, due to the presence of a jump; see Fig.~\ref{fig:single_switch}(b). For other values of applied force, the susceptibility is equal to zero both at zero temperature, when the system is ``frozen''   in one of the states, and at infinite temperature, when the mixing due to fluctuations dominates the bias introduced by the external force. Interestingly, it reaches a maximum at a finite value of \( \beta \); see Fig.~\ref{fig:SingleSwitchSusceptibility}(b).
The susceptibility \eqref{eq:SingleSwitchChi} is of interest because it is directly linked to the  equilibrium stiffness 
\begin{equation*}
	\kappa(\sigma,\beta) = \left[\frac{\partial\, \mean{y}}{\partial\sigma}\right]^{-1} = \left[1+\chi(\sigma,\beta)\right]^{-1}>0.
\end{equation*}
which has the Cauchy-Born  and the  fluctuations induced  contributions; the only difference with the behavior in a hard device \citep{Caruel:2016kw} is that the fluctuation part comes with plus sign, which makes the stiffness always positive.

\begin{figure}
	\centering
	\includegraphics[]{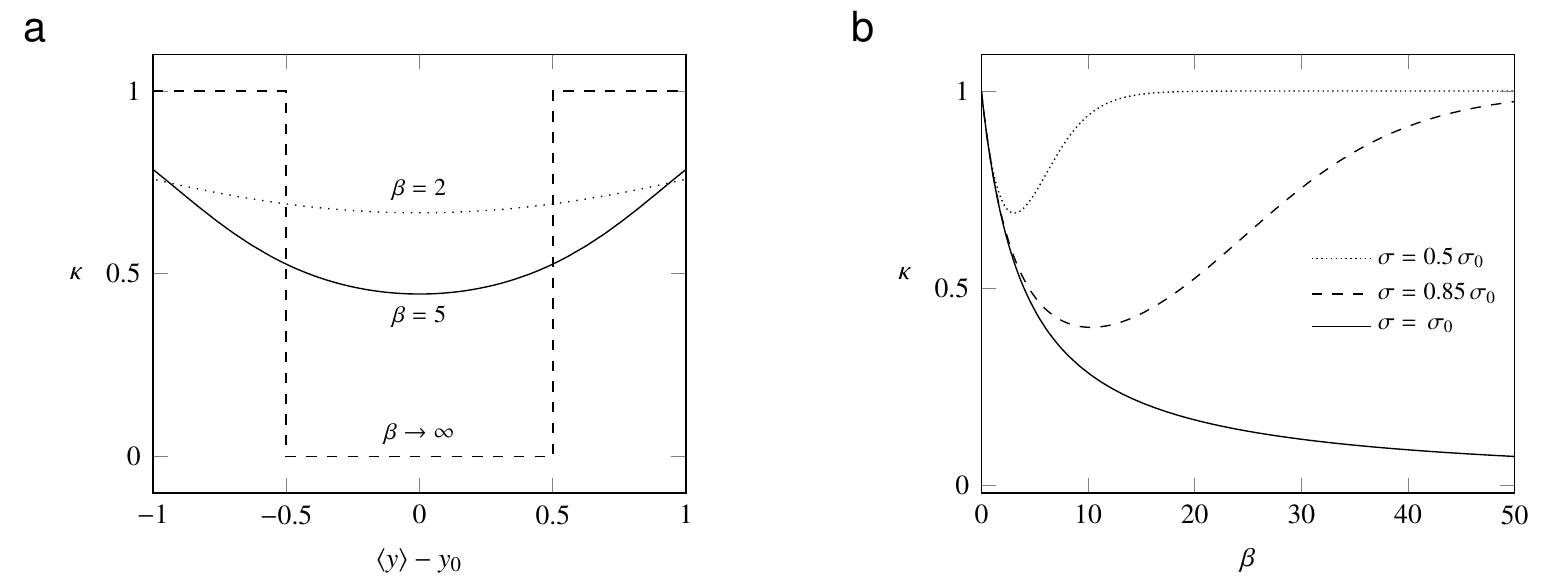}
	\caption{
	Equilibrium stiffness of a single digital switch represented as function of the loading (a), and as function of temperature (b).
	}
	\label{fig:SingleSwitchStiffness}
\end{figure}
The equilibrium stiffness $\kappa$  is represented as function of the external force and the temperature in Fig.~\ref{fig:SingleSwitchStiffness}.  Note that  for \( \beta\gg 1 \), the stiffness decreases with temperature (softening) while for \( \beta\to 0 \), it increases (hardening). The major difference with the hard device behavior occurs at \( \sigma=\sigma_0 \), where the susceptibility becomes infinite, which makes the stiffness decreasing towards 0 (instead of \( -\infty \) in a hard device) in the limit \( \beta\to\infty \).

Next we move to the discussion of the system of $N$ switching elements.  First, we compute the equilibrium susceptibility  
 \begin{equation}
	 \label{eq:susceptibility}
 	X(\sigma,\beta) = -\frac{1}{N}\frac{\partial}{\partial\sigma}\mean{p}(\sigma,\beta)= \beta \mean{\left[p-\mean{p}(\sigma,\beta)\right]^{2}}\geq 0.
 \end{equation}
 We define the rescaled susceptibility as \( \chi = NX \). In the thermodynamic limit, we have \( \mean{p} = p_{*}(\sigma,\beta) \) which gives
\begin{equation}
	\label{eq:x_infinity}
	\chi_{\infty}(\sigma,\beta)
	=-\frac{1}{N}\frac{\partial}{\partial\sigma}p_{*}(\sigma,\beta) 
	= \left\{\left[\beta p_{*}(\sigma,\beta)\left(1-p_{*}(\sigma,\beta)\right)\right]^{-1}-1\right\}^{-1}.
\end{equation}
 Expanding near the critical point and keeping  \( \sigma=\sigma_{0} \), we obtain
$
 	\chi_{\infty}(\sigma,\beta)\sim -  \left[1-\frac{4}{\beta}\right]^{-1},
$
which is the analog of the Curie--Weiss law in ferromagnetism \citep{Balian:2006wd}.  Notice that the susceptibility is proportional to the fluctuations, which take the form \( \mean{\left[p-\mean{p}\right]^{2}} \).
Expansion near \( \sigma=\sigma_0 \) for \( \beta=4 \) leads to 
$
	\chi_{\infty}(\sigma,\beta)\sim 4^{-1/3} \left[3(\sigma-\sigma_{0})\right]^{-2/3}.
$

 \begin{figure}[t]
 	\centering
 	\includegraphics[]{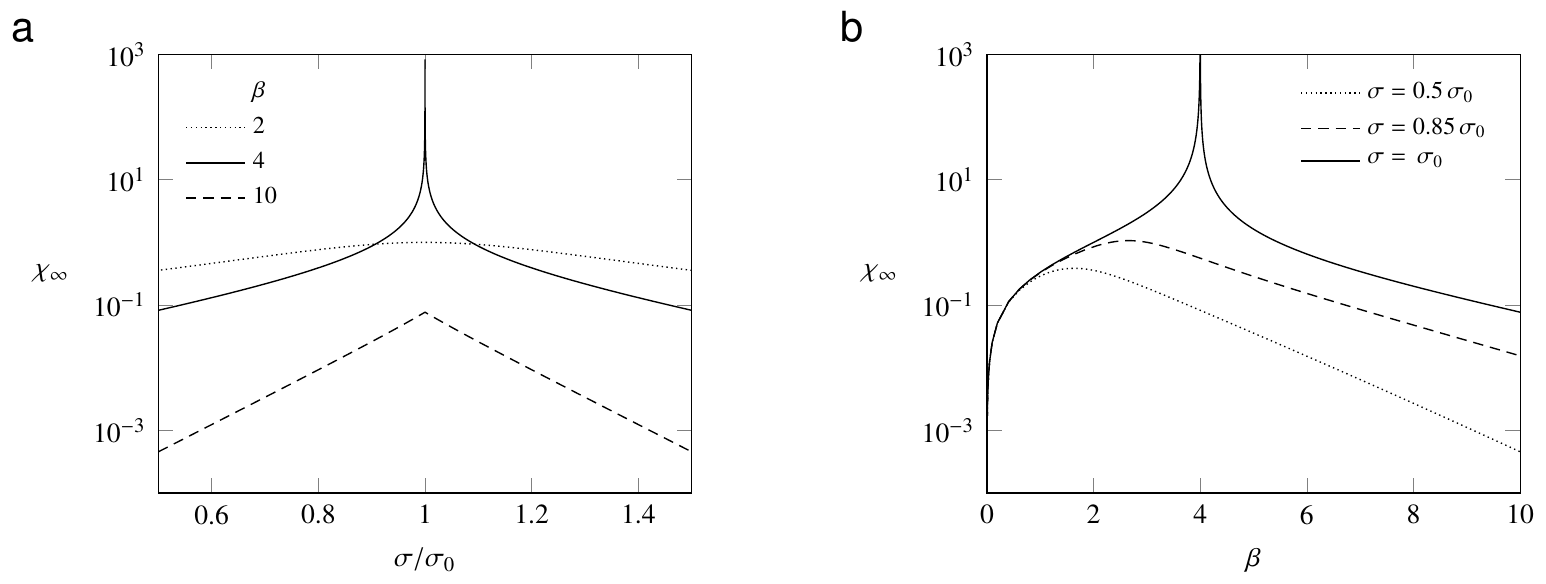}
 	\caption{
 	 Susceptibility of a cluster of bistable elements in the thermodynamic limit as function of the loading (a) and as function of temperature (b).
 	}
 	\label{fig:susceptibility}
 \end{figure}
The parameter dependence of the susceptibility $\chi_{\infty}$ is illustrated  in Fig.~\ref{fig:susceptibility}.
In agreement with what we learned from the study of a single  switching element (see Fig.~\ref{fig:SingleSwitchSusceptibility})   the susceptibility reaches its  maximum at \( \sigma=\sigma_0 \), and converges to 0 at both   \( \beta\to 0 \), and \( \beta\to\infty \). The main difference is due to the  presence of the phase transition occurring at \( \beta=4 \), and resulting in a  divergence  of susceptibility, which is behind large fluctuations observed in Fig.~\ref{fig:Average_configuration}(d).

\begin{figure}[htbp]
	\centering
	\includegraphics[]{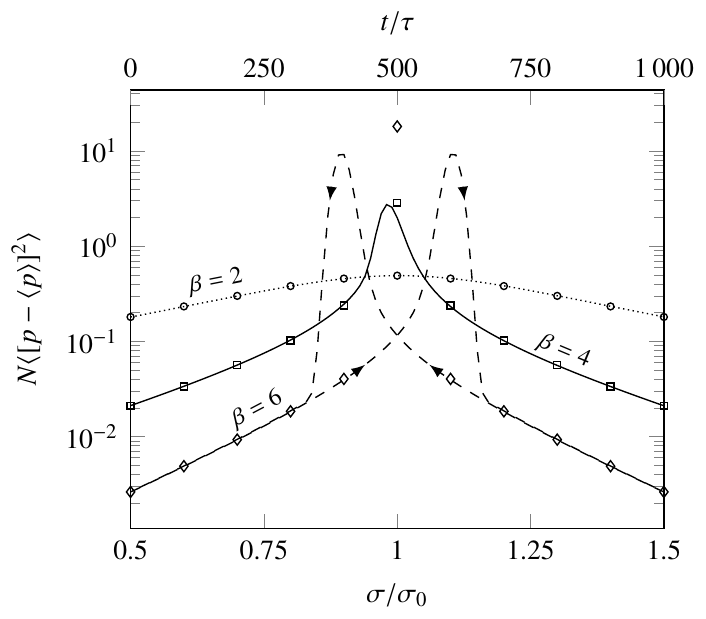}
	\caption{
	Evolution of the variance \( \mean{\left(p-\mean{p}\right)^{2}} \) obtained from the simulations shown in Fig.~\ref{fig:quasi_static_trajectories}. Dotted line, \( \beta=2 \); solid line, \( \beta=4 \); dashed lines, \( \beta=6 \). For \( \beta=6 \). The symbols on the lines show the variance obtained in thermal equilibrium from the distribution \eqref{eq:rho_bar}  with \( \beta=2 \) (circles), \( \beta =4 \)(squares) and \( \beta = 6 \) (diamonds).
	}
	\label{fig:quasi_static_variance}
\end{figure}

In Fig.~\ref{fig:quasi_static_variance} we show the time dependent variance of the evolving order parameter  \( p (t) \),  obtained from the simulations shown in Fig.~\ref{fig:quasi_static_trajectories}. 
The results obtained from both the solution of the master equation [\eqref{eq:master_equation}, lines] and the direct computation in thermal equilibrium (symbols) are shown. 
At supercritical temperatures and at the critical point, the equilibrium value fit with the numerical simulations well, which confirms  that the selected loading protocol  is essentially quasi-static; see Fig.~\ref{fig:quasi_static_variance}, dotted line.
However, for \( \beta=6 \), the equilibrium behavior---indicated by the diamond in Fig.~\ref{fig:quasi_static_variance}---is not recovered by the quasi-static simulations in the bi-stable region; see Fig.~\ref{fig:quasi_static_variance}, dashed line. Instead, kinetic simulations show  two peaks   corresponding to the collective  transitions observed when the system reaches the boundaries of the bistable domain during either loading or unloading.

By computing the derivative of the average displacement \eqref{eq:average_y} with respect to  \( \Sigma=N\sigma \), we obtain the expression for the equilibrium stiffness
\begin{equation*}
	K(\sigma,\beta)^{-1} = \frac{1}{N}\frac{\partial}{\partial\sigma}\mean{y}(\sigma,\beta)
	=\beta\mean{\left[y-\mean{y}(\sigma,\beta)\right]^{2}}\geq 0,
\end{equation*}
In the hard device case, the stiffness was shown to be  sign indefinite  ~\citep{Caruel:2016kw}, while here 
 \( K \) is   always positive which confirms that the two ensembles are not equivalent.
If  we denote by \( \kappa = K/N \), the stiffness per element, we obtain
\begin{equation*}
	\kappa(\sigma,\beta) = \left[1+\chi(\sigma,\beta)\right]^{-1},
\end{equation*}
where the susceptibility \( \chi \) was obtained  earlier, see  Eq.~\eqref{eq:susceptibility}. 
In the thermodynamic limit, 
\(
	\kappa_{\infty}(\sigma,\beta) = \left[1+\chi_{\infty}(\sigma,\beta)\right]^{-1}
\)
where \( \chi_{\infty} \) is given by Eq.~\eqref{eq:x_infinity}.
  
\begin{figure}
	\centering
	\includegraphics[]{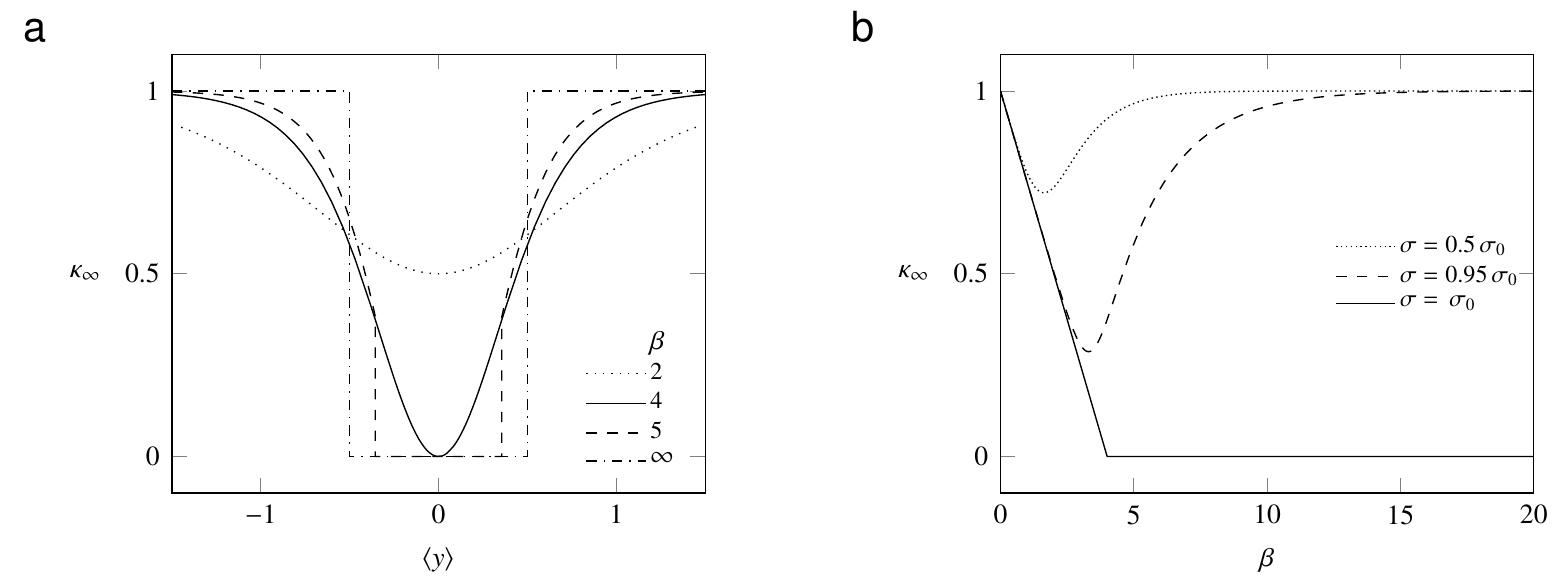}
	\caption{
	Stiffness of a cluster in the thermodynamic limit. (a) As function of the applied load. (b) As function of the temperature
	}
	\label{fig:cluster_stiffness}
\end{figure}

The dependence of the stiffness on the applied load and the temperature is illustrated  in Fig.~\ref{fig:cluster_stiffness}.
For \( \beta<\beta_c \), the behavior is similar to the case of a single switching element, see Fig.~\ref{fig:SingleSwitchStiffness}: the stiffness is positive, reaches a    temperature independent minimum at \( \sigma=\sigma_0 \) [see (a)] and is equal to 1 at both infinite and zero temperatures [see (b)].
For \( \beta>\beta_c \), the two behaviors are rather different.
The stiffness of the bundle shows a plateau at a range of temperatures  [see Fig.~\ref{fig:cluster_stiffness}(a)], while in the case of a single  element a plateau was obtained only at zero temperature.

\section{Thermal behavior} 
\label{sec:thermal_behavior}

In addition to the mechanical properties of our ``device'', it will be of interest to study their thermal properties.   First, we present the study the equilibrium thermal behavior of a single switching element. From the partition function \eqref{eq:single_switch_partition_function}, we obtain the expression of the equilibrium entropy,
\begin{equation}
	\label{eq:single_switch_entropy}
	s(\sigma,\beta) = -\beta \frac{\partial}{\partial\beta}\log\left[Z\right] + \log[Z] = \frac{1}{2} + \frac{1}{2}\log\left[\frac{2\pi}{\beta}\right] + \log\left\{2\cosh\left[\frac{\beta}{2}\left(\sigma-\sigma_0\right)\right]\right\} - \frac{\beta}{2}(\sigma-\sigma_0)\tanh\left[\frac{\beta}{2}(\sigma-\sigma_0)\right].
\end{equation}
\begin{figure}[t]
	\centering
	\includegraphics[]{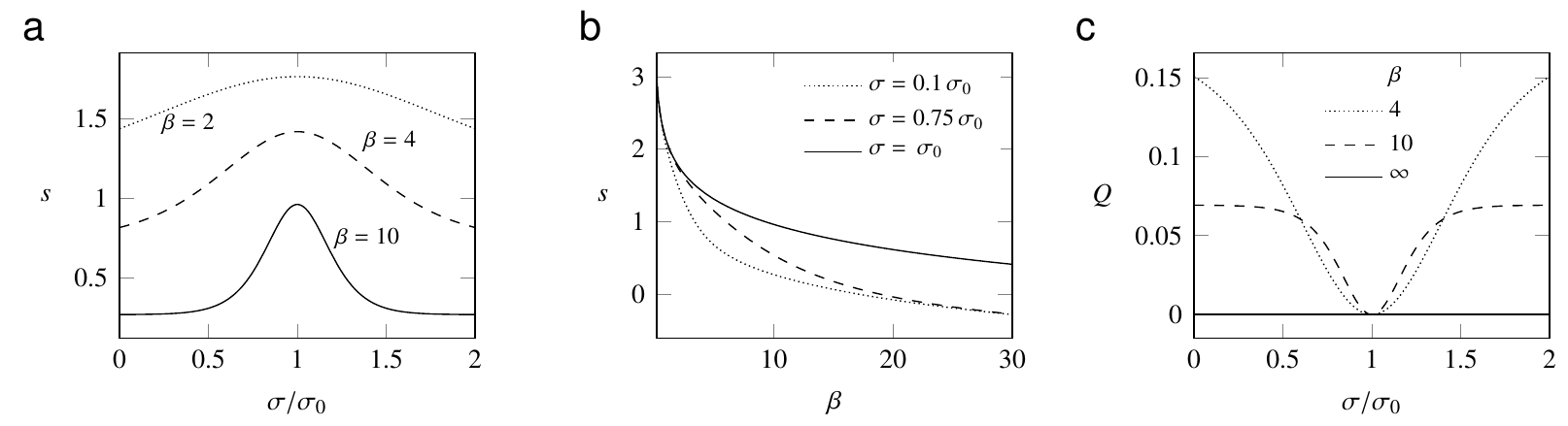}
	\caption{
	Entropy of a single digital switch as function of the external force at different temperatures (a) and as function of temperature for different external forces (b). (b) Heat release in response to a load change from \( \sigma=\sigma_0 \) for different temperatures.
	}
	\label{fig:SingleSwitchEntropy}
\end{figure}
The entropy $s$ is represented as a function of the external load and   temperature in Fig.~\ref{fig:SingleSwitchEntropy}[(a) and (b)]. It reaches a maximum at \( \sigma=\sigma_0 \), and decreases to a constant (temperature-dependent) value away from this point. The difference between the maximum and the minimum entropy reaches a constant value equal to \( \log(2) \) at low temperatures, see Fig.~\ref{fig:SingleSwitchEntropy}(b). 

The dependence of the entropy \eqref{eq:single_switch_entropy} on  external load characterizes the heat release  \( Q\left(\sigma,\beta\right) = -(1/\beta)\Delta s(\sigma,\beta) \), where \( \Delta s(\sigma,\beta) = s(\sigma,\beta) - s(\sigma_{\mbox{\tiny in}},\beta) \) is the entropy increment starting from the initial state with \( \sigma_{\mbox{\tiny in}} \).  The heat release in response to loading  from the point \( \sigma=\sigma_0 \) is represented for different temperatures in Fig.~\ref{fig:SingleSwitchEntropy}(c).  
\begin{figure}[t]
	\centering
	\includegraphics[]{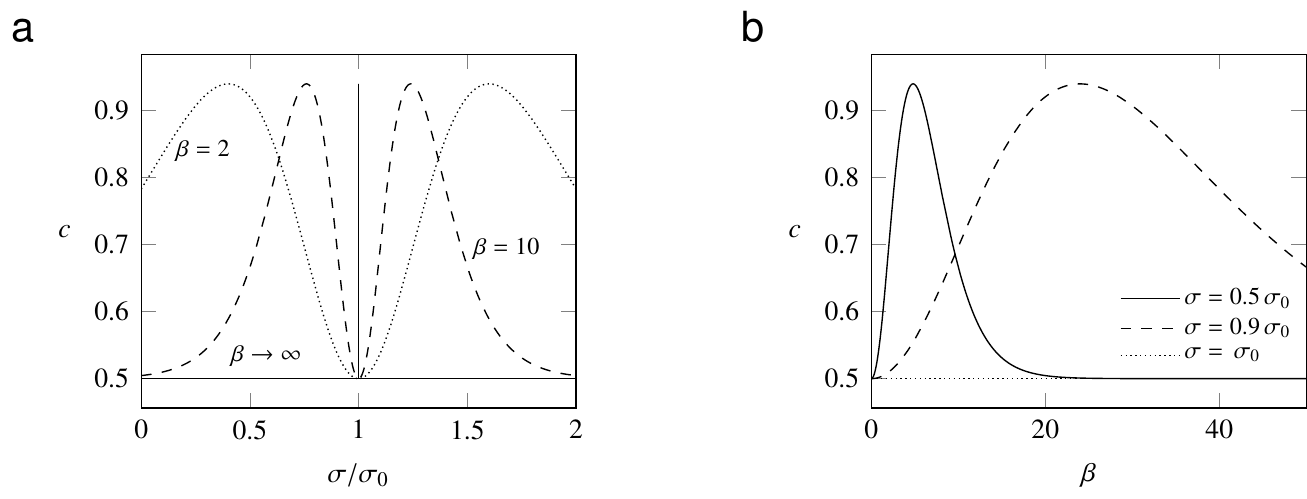}
	\caption{
	Specific heat of a single digital switch as function of the loading (a) and as function of temperature (b).
	}
	\label{fig:SingleSwitchSpecificHeat}
\end{figure}
The specific heat \( c \) at constant load   
\begin{equation*}
	c(\sigma,\beta) = -\beta\frac{\partial}{\partial \beta}s(\sigma,\beta) = \frac{1}{2} + \left\{\frac{\beta}{2}\left(\sigma-\sigma_0\right)\,\sech\left[\frac{\beta}{2}\left(\sigma-\sigma_0\right)\right]\right\}^{2},
\end{equation*}
 is illustrated in Fig.~\ref{fig:SingleSwitchSpecificHeat}.
We see that specific heat is equal to zero at \( \sigma_0 \) making this state insensitive (robust) to temperature changes. Similarly at large value of the applied force, the mechanical bias of the system dominates the fluctuations, which again makes the system temperature insensitive. 

\begin{figure}[htbp]
	\centering
	\includegraphics[]{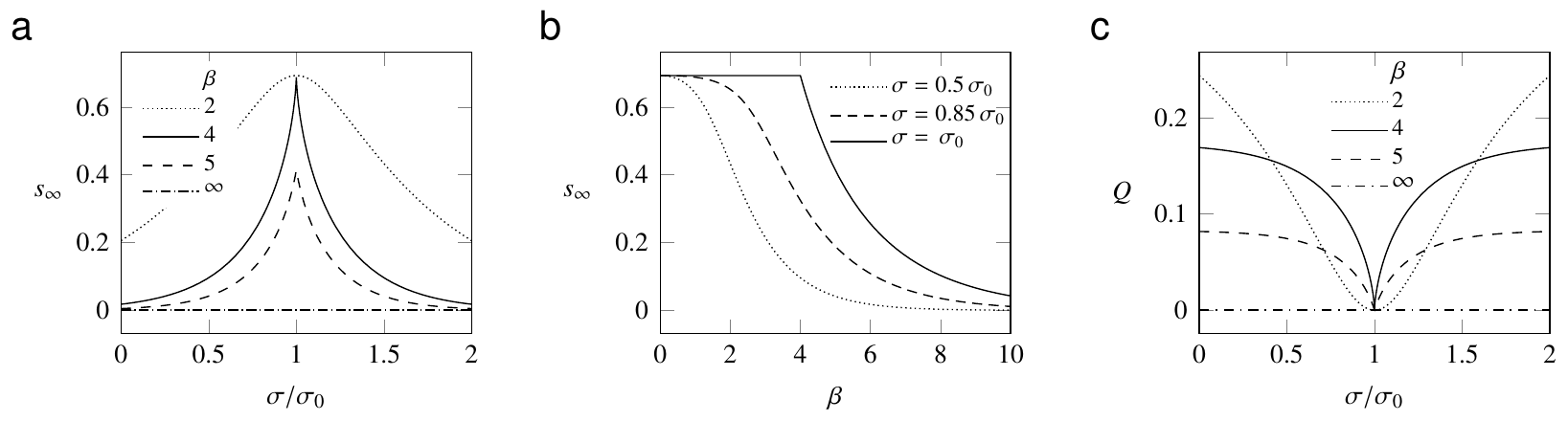}
	\caption{
	Entropy as function of the loading (a) and as function of temperature (b) in the thermodynamics limit. (c) Heat release for different temperatures with the reference state at \( \sigma=\sigma_0 \).
	}
	\label{fig:entropy}
\end{figure}

Next we study  a bundle of $N$ switching elements.   We first define   the  entropy per element  $s $  
\begin{equation*}
	Ns(\sigma,\beta) = -\beta\frac{\partial}{\partial\beta}\log\left[Z(\sigma,\beta )\right] + \log\left[Z(\sigma,\beta)\right].
\end{equation*}
The function $s(\sigma,\beta)$ can be also written as   \( s = \beta\mean{e} - \beta g \), with the  average  internal energy is 
\begin{equation}
	\label{eq:mean_e}
		\mean{e}(\sigma,\beta) 
		 =  \sum_{p} \bar{w}(p;\sigma)\,\bar{\rho}(p;\sigma,\beta) + 1/\left(2\,\beta\right).
\end{equation}
In the thermodynamic limit \eqref{eq:mean_e} becomes
$
		\mean{e}(\sigma,\beta)_{\infty} (\sigma,\beta) 		=  w\left[p_{*}(\sigma,\beta);\sigma,\beta\right],
$
and  
$
		s_{\infty}(\sigma,\beta)= s_{\infty}\left[p_{*}(\sigma,\beta)\right].
$
The dependence of the entropy on temperature and loading is illustrated in Fig.~\ref{fig:entropy}. The entropy is maximized at \( \sigma=\sigma_{0} \), where  \( \mean{p}=1/2 \), and develops a singularity at this point below the critical temperature; see Fig.~\ref{fig:entropy}(a), dashed line. It  is always an increasing function of temperature for \( \sigma\neq\sigma_0 \), except when \( \sigma=\sigma_0 \), where it is independent of temperature for \( \beta<\beta_{c} \), and is a  decreasing function of temperature  for \( \beta>\beta_c \); see Fig.~\ref{fig:entropy}(a).
 
The variations of entropy  as one goes from one equilibrium state to another can be linked  to the heat exchange with the surrounding thermostat :
  \( Q = -\beta^{-1}\Delta s(\sigma,\beta) \), where \( \Delta s (\sigma,\beta) = s(\sigma,\beta) - s(\sigma_{\text{\tiny in}},\beta) \).  
Our Fig.~\ref{fig:entropy}(c) illustrates the heat release starting from the initial equilibrium state at \( \sigma_{\text{\tiny in}} = \sigma_0 \). Below the critical point (dotted line, \( \beta=2 \)), the heat release is negligible in the vicinity of the initial state as expected from the smooth dependence of the entropy on applied force; see Fig.~\ref{fig:entropy}(a). Above the critical point, the entropy develops a cusp and therefore the heat is released abruptly even for a small external perturbation; see Fig.~\ref{fig:entropy}[(c), solid and dashed lines].

\begin{figure}
	\centering
	\includegraphics[]{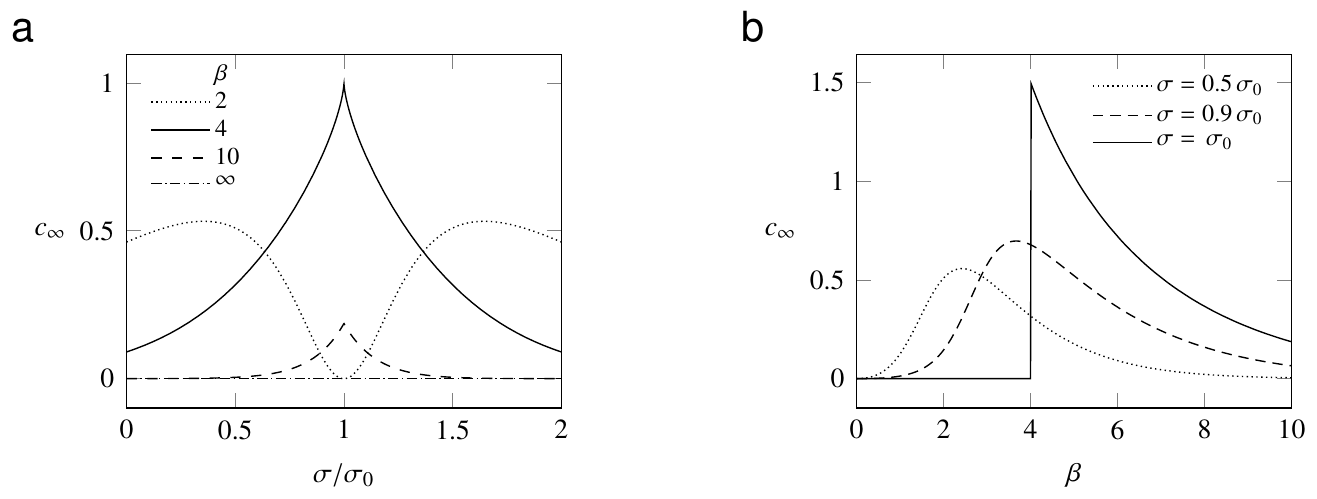}
	\caption{Specific heat in the thermodynamic limit as function of the loading (a) and as function of temperature (b). Here, \( v_0=1 \).
	}
	\label{fig:specific_heat}
\end{figure}

The fact that the capacity of the system to exchange heat with the thermostat is maximal at the critical point can also be seen  from the   specific heat 
\begin{equation*}
		c(\sigma,\beta)
		=-\beta \frac{\partial s}{\partial \beta} 
		= \frac{1}{2N} + N\beta^{2}\mean{\left[w(p;\sigma,\beta) - \mean{w}(\sigma,\beta)\right]^{2}}.
\end{equation*}
In the thermodynamic limit  
\begin{equation*}
	c_{\infty}(\sigma,\beta) =  \frac{1}{\beta}\left\{\log\left[\frac{p_{*}}{1-p_{*}}\right]\right\}^{2}\chi_{\infty}(\sigma,\beta),
\end{equation*}
where \( X_{\infty} \) is given by Eq.~\ref{eq:x_infinity}.
Expanding for  \( \beta>4 \) at  \( \sigma=\sigma_0 \), we obtain
$
	c_{\infty}(\sigma,\beta)\sim\frac{3}{2}\left[1-\frac{5}{8}\left(\beta-4\right)\right],
$
Similar expansion near \( \sigma=\sigma_0 \) for \( \beta=4 \), leads to
\( 
	c_{\infty}(\sigma,\beta) = 1-3^{-1/3}\left(2\left|\sigma-\sigma_0\right|\right)^{2/3}.
 \)
These results are illustrated in Fig.~\ref{fig:specific_heat}.  

\begin{figure*}
	\centering
	\includegraphics[]{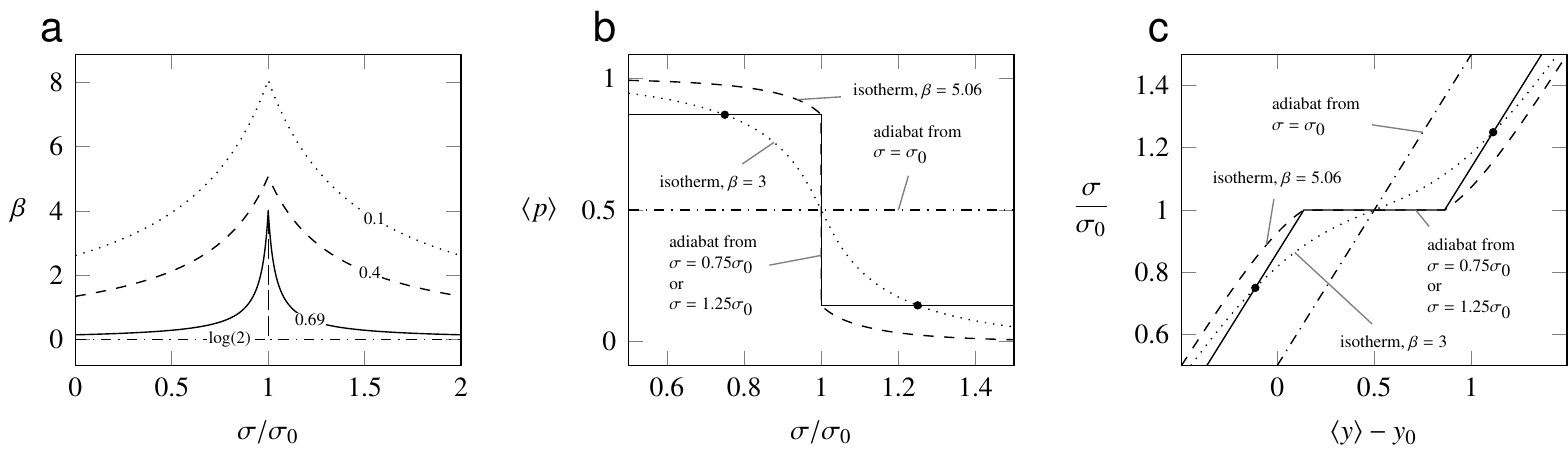}
	\caption{
	Adiabatic response. (a) isoentropes shown for different values of entropy. (b) evolution of \( \mean{p} \) along adiabats in the thermodynamic limit. Dot-dashed, adiabatic response obtained starting from thermal equilibrium at \( \sigma=\sigma_0 \); Dotted, isotherm corresponding to \( \beta=3 \). Solid, adiabatic response obtained when loading the system from the equilibrium state at \( \sigma=0.75\sigma_0 \) and \( \sigma=1.25\sigma_0 \) with \( \beta=3 \) (see the black dots). As the load changes, the value \( |p-1/2| \) is kept constant by a temperature change up to \( \beta=5.06 \) at \( \sigma=\sigma_0 \). The isotherm corresponding to \( \beta=5.06 \) is represented by the dashed line. Here \( v_0=1 \).
	}
	\label{fig:adiabats}
\end{figure*}

When the  loading is fast at  the timescale of heat exchange with the thermostat,  the mechanical response becomes adiabatic.  
To assess  the associated  temperature variations, we recall that the entropy depends only   on \( |p-1/2| \).
Therefore, for an initial state with entropy \( s_{\mbox{\tiny in}} \) corresponding to   \( p=p_{*\mbox{\tiny in}} \) we obtain along the adiabat
\begin{equation}
	\beta_{\mbox{\tiny AD}}(\sigma) = 
	\left[
		\sigma-\sigma_0+\mbox{sign}(\sigma-\sigma_0)\left|\frac{1}{2}-p_{*\mbox{\tiny in}}\right|\right
	]^{-1}
	\log\left[\frac{\frac{1}{2}+\mbox{sign}(\sigma-\sigma_0)|\frac{1}{2}-p_{*\mbox{\tiny in}}|}{\frac{1}{2}-\mbox{sign}(\sigma-\sigma_0)|\frac{1}{2}-p_{*\mbox{\tiny in}}|}\right].
\end{equation}
Several typical adiabats of this type (isoentropes)  are presented in Fig.~\ref{fig:adiabats} (a). If the system is prepared in an initial equilibrium state at \( \sigma>\sigma_0 \), and the load is released, the conservation of the initial entropy leads to a decrease in temperature up a finite value of \( \beta \), but then the temperature rises again as the load continues to decrease  for  \( \sigma<\sigma_0 \).  

The reason why the temperature drops to a finite value during the unloading  becomes clear when we examine  Fig.~\ref{fig:adiabats} (b),  showing the evolution of \( \mean{p} \) along the adiabats. Suppose the system is prepared in thermal equilibrium at \( \beta=3 \), and \( \sigma=1.25\sigma_0 \) [see Fig.~\ref{fig:adiabats}(b), black dot where \( p_{\mbox\tiny in}\approx0.89 \)], and then the load is dropped to a value below \( \sigma_0 \).
 First, the temperature decreases as the system has to maintain the initial value \( \mean{p}  =  p_{\mbox\tiny in} \), and this process continues till the applied load reaches the value \( \sigma_0 \). The value of \( \beta_{\mbox\tiny AD} \), ensuring that \( \mean{p}(\sigma_0,\beta_{\mbox\tiny AD}) = p_{\mbox\tiny in} \), is about \( 5.06 \).
The isotherm corresponding to \( \beta=5.06 \) is represented by the dashed line in Fig.~\ref{fig:adiabats}(b), and we see that the adiabatic trajectory (solid line) intercept the isotherm  at the jump point. As the load is decreased further, the  temperature starts to increase from the value \( \beta=5.06 \) as the system has to maintain   the new value of \( \mean{p}  = 1-p_{\mbox\tiny in} \).

\section{Boundaries of the bistable domain} 
\label{sec:boundaries_of_the_bistable_domain}

The critical points of \( \bar{g}_{\infty} \) are solutions of
\( \frac{\partial}{\partial p}\bar{g}_{\infty}=0 \), namely of
\begin{equation}
	\label{eq:d_g_bar_p_infinity_d_p=0}
	\sigma-\sigma_0=p-\frac{1}{2}- \frac{1}{\beta}\ln\left[\frac{p}{1-p}\right],
\end{equation}
for which the number of solutions depends on the variations of the right hand side. We have,
\[
	\frac{\partial^{2}\bar{g}_{\infty}}{\partial p^{2}} = -1 + \left[\beta p(1-p)\right]^{-1}
\]
which when \( \beta>4 \) is equal to zero for \( p=p_{+}(\beta)\)  and \( p=p_{-}(\beta) \) with
\begin{equation}\label{eq:sd_p_+_p_-}
	\begin{split}
		p_{+}(\beta) &= (1+\sqrt{1-4\beta^{-1}})/2\\
		p_{-}(\beta) &= (1-\sqrt{1-4\beta^{-1}})/2.
	\end{split}
\end{equation}
By inserting these values in Eq.~\ref{eq:d_g_bar_p_infinity_d_p=0} we obtain,  \begin{equation}
	\label{eq:sigma_+_sigma_-}
	\begin{split}
		\sigma_{+}(\beta) &= \sigma_{0} +
		\frac{1}{2}\sqrt{1-\frac{4}{\beta}}
		-\frac{1}{\beta}\log\left[
		\frac{\sqrt{\beta}+\sqrt{\beta-4}}
		{\sqrt{\beta}-\sqrt{\beta-4}}\right]\\
		\sigma_{-}(\beta) &= \sigma_{0} - \frac{1}{2}\sqrt{1-\frac{4}{\beta}}+
		\frac{1}{\beta}\log\left[
		\frac{\sqrt{\beta}+\sqrt{\beta-4}}
		{\sqrt{\beta}-\sqrt{\beta-4}}\right]
	\end{split}
\end{equation}
Hence we find that the system has 3 critical points when \( \sigma_{-}(\beta)<\sigma<\sigma_{+}(\beta) \). In the limit \( \beta\to\infty \) we obtain that \( \sigma_{+}=v_0+1/2 \) and to \( \sigma_{-}=v_{0}-1/2 \), when \( \beta\to\infty \), respectively; see \cite{Caruel:2015im}.

\section*{\refname}
\bibliography{biblio}

\begin{thebibliography}{52}
\expandafter\ifx\csname natexlab\endcsname\relax\def\natexlab#1{#1}\fi
\expandafter\ifx\csname url\endcsname\relax
  \def\url#1{\texttt{#1}}\fi
\expandafter\ifx\csname urlprefix\endcsname\relax\def\urlprefix{URL }\fi

\bibitem[{Achenbach et~al.(1986)Achenbach, Atanackovic, and
  M{\"u}ller}]{Achenbach:1986kh}
Achenbach, M., Atanackovic, T., M{\"u}ller, I., 1986. {A model for memory
  alloys in plane strain}. International Journal of Solids and Structures
  22~(2), 171--193.

\bibitem[{Balian(2006)}]{Balian:2006wd}
Balian, R., 2006. {From Microphysics to Macrophysics}. Methods and Applications
  of Statistical Physics. Springer Science {\&} Business Media.

\bibitem[{Barr{\'e} et~al.(2002)Barr{\'e}, Mukamel, and Ruffo}]{Barre:2002ck}
Barr{\'e}, J., Mukamel, D., Ruffo, S., 2002. {Ensemble Inequivalence in
  Mean-Field Models of Magnetism}. In: Dynamics and Thermodynamics of Systems
  with Long-Range Interactions. Springer Berlin Heidelberg, Berlin, Heidelberg,
  pp. 45--67.

\bibitem[{Bell(1978)}]{Bell_1978}
Bell, G., 1978. {Model for the Specific Adhesion of Cells to Cells}. Science
  200.

\bibitem[{Bell and Terentjev(2017)}]{Bell:2017ct}
Bell, S., Terentjev, E.~M., Jun. 2017. {Focal Adhesion Kinase: The Reversible
  Molecular Mechanosensor.} Biophys. J. 112~(11), 2439--2450.

\bibitem[{Benichou and Givli(2015)}]{Benichou:2015je}
Benichou, I., Givli, S., Mar. 2015. {Rate dependent response of nanoscale
  structures having a multiwell energy landscape.} Phys. Rev. Lett. 114~(9),
  095504.

\bibitem[{Benichou et~al.(2016)Benichou, Zhang, Dudko, and
  Givli}]{Benichou:2016br}
Benichou, I., Zhang, Y., Dudko, O.~K., Givli, S., Oct. 2016. {The rate
  dependent response of a bistable chain at finite temperature}. J. Mech. Phys.
  Solids 95, 44--63.

\bibitem[{Bormuth et~al.(2014)Bormuth, Barral, Joanny, J{\"u}licher, and
  Martin}]{Bormuth:2014hh}
Bormuth, V., Barral, J., Joanny, J.~F., J{\"u}licher, F., Martin, P., May 2014.
  {Transduction channels' gating can control friction on vibrating hair-cell
  bundles in the ear}. Proc. Natl. Acad. Sci. U.S.A. 111~(20), 7185--7190.

\bibitem[{B{\"u}ckmann et~al.(2015)B{\"u}ckmann, Kadic, Schittny, and
  Wegener}]{Buckmann:2015cr}
B{\"u}ckmann, T., Kadic, M., Schittny, R., Wegener, M., Apr. 2015. {Mechanical
  cloak design by direct lattice transformation.} Proc. Natl. Acad. Sci. U.S.A.
  112~(16), 4930--4934.

\bibitem[{Caruel et~al.(2015)Caruel, Allain, and Truskinovsky}]{Caruel:2015im}
Caruel, M., Allain, J.-M., Truskinovsky, L., Mar. 2015. {Mechanics of
  collective unfolding}. J. Mech. Phys. Solids 76, 237--259.

\bibitem[{Caruel and Truskinovsky(2016)}]{Caruel:2016kw}
Caruel, M., Truskinovsky, L., Jun. 2016. {Statistical mechanics of the
  Huxley-Simmons model}. Phys. Rev. E 93~(6), 062407.

\bibitem[{Cho et~al.(2016)Cho, Weaver, P{\"o}selt, and Boyce}]{Cho:2016kh}
Cho, H., Weaver, J.~C., P{\"o}selt, E., Boyce, M.~C., 2016. {Engineering the
  Mechanics of Heterogeneous Soft Crystals}. Advanced Functional Materials
  26~(38), 6938--6949.

\bibitem[{Desai and Zwanzig(1978)}]{Desai_1978}
Desai, R.~C., Zwanzig, R., 1978. {Statistical mechanics of a nonlinear
  stochastic model}. J. Stat. Phys. 19, 1--24.

\bibitem[{Efendiev and Truskinovsky(2010)}]{Efendiev_2010}
Efendiev, Y.~R., Truskinovsky, L., Sep. 2010. {Thermalization of a driven
  bi-stable FPU chain}. Continuum Mech. Therm. 22~(6-8), 679--698.

\bibitem[{Flood et~al.(2004)Flood, Stoddart, Steuerman, and
  Heath}]{Flood:2004kr}
Flood, A.~H., Stoddart, J.~F., Steuerman, D.~W., Heath, J.~R., Dec. 2004.
  {Whence Molecular Electronics?} Science 306~(5704), 2055--2056.

\bibitem[{Florijn et~al.(2014)Florijn, Coulais, and van Hecke}]{Florijn:2014fw}
Florijn, B., Coulais, C., van Hecke, M., Oct. 2014. {Programmable mechanical
  metamaterials.} Phys. Rev. Lett. 113~(17), 175503.

\bibitem[{Frenzel et~al.(2016)Frenzel, Findeisen, Kadic, Gumbsch, and
  Wegener}]{Frenzel:2016hg}
Frenzel, T., Findeisen, C., Kadic, M., Gumbsch, P., Wegener, M., Jul. 2016.
  {Tailored Buckling Microlattices as Reusable Light-Weight Shock Absorbers.}
  Adv. Mater. Weinheim 28~(28), 5865--5870.

\bibitem[{Gardiner(2004)}]{Gardiner_2004}
Gardiner, C., 2004. Handbook of stochastic methods for physics chemistry and
  the natural sciences, 3rd Edition. Springer.

\bibitem[{Haghpanah et~al.(2016)Haghpanah, Salari-Sharif, Pourrajab, Hopkins,
  and Valdevit}]{Haghpanah:2016ih}
Haghpanah, B., Salari-Sharif, L., Pourrajab, P., Hopkins, J., Valdevit, L.,
  Sep. 2016. {Architected Materials: Multistable Shape-Reconfigurable
  Architected Materials (Adv. Mater. 36/2016)}. Advanced Materials 28~(36),
  8065--8065.

\bibitem[{Halg(1990)}]{Halg:1990fq}
Halg, B., Oct. 1990. {On a micro-electro-mechanical nonvolatile memory cell}.
  IEEE Transactions on Electron Devices 37~(10), 2230--2236.

\bibitem[{Harne and Wang(2013)}]{Harne:2013if}
Harne, R.~L., Wang, K.~W., Feb. 2013. {A review of the recent research on
  vibration energy harvesting via bistable systems}. Smart Mater. Struct.
  22~(2), 023001.

\bibitem[{{Harne, R L} et~al.(2016){Harne, R L}, {Wu, Z}, and {Wang, K
  W}}]{Harne:2016gg}
{Harne, R L}, {Wu, Z}, {Wang, K W}, Feb. 2016. {Designing and Harnessing the
  Metastable States of a Modular Metastructure for Programmable Mechanical
  Properties Adaptation}. J. Mech. Des 138~(2), 021402.

\bibitem[{Hill(1974)}]{Hill:1974ks}
Hill, T.~L., Jan. 1974. {Theoretical formalism for the sliding filament model
  of contraction of striated muscle Part I}. Prog. Biophys. Molec. Biol. 28,
  267--340.

\bibitem[{Hill(1976)}]{Hill:1976gf}
Hill, T.~L., Jan. 1976. {Theoretical formalism for the sliding filament model
  of contraction of striated muscle part II}. Prog. Biophys. Molec. Biol. 29,
  105--159.

\bibitem[{Howard and Hudspeth(1988)}]{Howard:1988uu}
Howard, J., Hudspeth, A.~J., May 1988. {Compliance of the hair bundle
  associated with gating of mechanoelectrical transduction channels in the
  bullfrog's saccular hair cell.} Neuron 1~(3), 189--199.

\bibitem[{Huxley and Simmons(1971)}]{Huxley_1971}
Huxley, A.~F., Simmons, R.~M., Oct. 1971. {Proposed Mechanism of Force
  Generation in Striated Muscle}. Nature 233, 533--538.

\bibitem[{Huxley and Tideswell(1996)}]{Huxley_1996}
Huxley, A.~F., Tideswell, S., Aug. 1996. {Filament compliance and tension
  transients in muscle}. J. Muscle Re. Cell M. 17, 507--511.

\bibitem[{Kometani and Shimizu(1975)}]{Kometani_1975}
Kometani, K., Shimizu, H., 1975. {A study of self-organizing processes of
  nonlinear stochastic variables}. J. Stat. Phys. 13, 473--490.

\bibitem[{Lebowitz and Penrose(1966)}]{Lebowitz:1966dj}
Lebowitz, J.~L., Penrose, O., Jan. 1966. {Rigorous Treatment of the Van Der
  Waals-Maxwell Theory of the Liquid-Vapor Transition}. Journal of Mathematical
  Physics 7~(1), 98--113.

\bibitem[{Li and Gao(2016)}]{Li:2016hs}
Li, X., Gao, H., Apr. 2016. {Mechanical metamaterials: Smaller and stronger.}
  Nat Mater 15~(4), 373--374.

\bibitem[{Linari et~al.(2010)Linari, Caremani, and Lombardi}]{Linari_2010a}
Linari, M., Caremani, M., Lombardi, V., Jan. 2010. {A kinetic model that
  explains the effect of inorganic phosphate on the mechanics and energetics of
  isometric contraction of fast skeletal muscle}. P Roy. Soc. Lond. B Bio.
  277~(270), 19--27.

\bibitem[{Martin et~al.(2000)Martin, Mehta, and Hudspeth}]{Martin_2000}
Martin, P., Mehta, A., Hudspeth, A., Oct. 2000. {Negative hair-bundle stiffness
  betrays a mechanism for mechanical amplification by the hair cell}. Proc.
  Natl. Acad. Sci. USA 97~(22), 12026--12031.

\bibitem[{Nadkarni et~al.(2016)Nadkarni, Arrieta, Chong, Kochmann, and
  Daraio}]{Nadkarni:2016jn}
Nadkarni, N., Arrieta, A.~F., Chong, C., Kochmann, D.~M., Daraio, C., Jun.
  2016. {Unidirectional Transition Waves in Bistable Lattices.} Phys. Rev.
  Lett. 116~(24), 244501.

\bibitem[{Nicolaou and Motter(2012)}]{Nicolaou:2012cf}
Nicolaou, Z.~G., Motter, A.~E., May 2012. {Mechanical metamaterials with
  negative compressibility transitions}. Nat Mater 11~(7), 608--613.

\bibitem[{Offer and Ranatunga(2016)}]{Offer:2016fn}
Offer, G., Ranatunga, K.~W., Nov. 2016. {Reinterpretation of the Tension
  Response of Muscle to Stretches and Releases.} Biophys. J. 111~(9),
  2000--2010.

\bibitem[{Ollier et~al.(1995)Ollier, Labeye, and Revol}]{Ollier:1995hi}
Ollier, E., Labeye, P., Revol, F., Nov. 1995. {Micro-opto mechanical switch
  integrated on silicon}. Electronics Letters 31~(23), 2003--2005.

\bibitem[{Paulose et~al.(2015)Paulose, Meeussen, and Vitelli}]{Paulose:2015hda}
Paulose, J., Meeussen, A.~S., Vitelli, V., Jun. 2015. {Selective buckling via
  states of self-stress in topological metamaterials.} Proc. Natl. Acad. Sci.
  U.S.A. 112~(25), 7639--7644.

\bibitem[{Piazzesi and Lombardi(1995)}]{Piazzesi_1995}
Piazzesi, G., Lombardi, V., May 1995. {A cross-bridge model that is able to
  explain mechanical and energetic properties of shortening muscle}. Biophys.
  J. 68~(5), 1966--1979.

\bibitem[{Rafsanjani et~al.(2015)Rafsanjani, Akbarzadeh, and
  Pasini}]{Rafsanjani:2015fh}
Rafsanjani, A., Akbarzadeh, A., Pasini, D., 2015. {Snapping mechanical
  metamaterials under tension}. Advanced Materials 27~(39), 5930--5930.

\bibitem[{Receveur et~al.(2005)Receveur, Marxer, Woering, Larik, and
  de~Rooij}]{Receveur:2005eu}
Receveur, R. A.~M., Marxer, C.~R., Woering, R., Larik, V. C. M.~H., de~Rooij,
  N.~F., Sep. 2005. {Laterally moving bistable MEMS DC switch for biomedical
  applications}. J. Microelectromech. Syst. 14~(5), 1089--1098.

\bibitem[{Reconditi et~al.(2004)Reconditi, Linari, Lucii, Stewart, Sun,
  Boesecke, Narayanan, Fischetti, Irving, Piazzesi, Irving, and
  Lombardi}]{Reconditi_2004}
Reconditi, M., Linari, M., Lucii, L., Stewart, A., Sun, Y., Boesecke, P.,
  Narayanan, T., Fischetti, R., Irving, T., Piazzesi, G., Irving, M., Lombardi,
  V., Apr. 2004. {The myosin motor in muscle generates a smaller and slower
  working stroke at higher load}. Nature 428~(6982), 578--581.

\bibitem[{Shan et~al.(2015)Shan, Kang, Raney, Wang, Fang, Candido, Lewis, and
  Bertoldi}]{Shan:2015fz}
Shan, S., Kang, S.~H., Raney, J.~R., Wang, P., Fang, L., Candido, F., Lewis,
  J.~A., Bertoldi, K., Aug. 2015. {Multistable Architected Materials for
  Trapping Elastic Strain Energy.} Adv. Mater. Weinheim 27~(29), 4296--4301.

\bibitem[{Sheshka et~al.(2016)Sheshka, Recho, and
  Truskinovsky}]{Sheshka:2016dm}
Sheshka, R., Recho, P., Truskinovsky, L., May 2016. {Rigidity generation by
  nonthermal fluctuations}. Phys. Rev. E 93~(5), 052604--14.

\bibitem[{Shiino(1987)}]{Shiino:1987tc}
Shiino, M., Sep. 1987. {Dynamical behavior of stochastic systems of infinitely
  many coupled nonlinear oscillators exhibiting phase transitions of mean-field
  type: H theorem on asymptotic approach to equilibrium and critical slowing
  down of order-parameter fluctuations.} Phys Rev A Gen Phys 36~(5),
  2393--2412.

\bibitem[{Smith et~al.(2008)Smith, Geeves, Sleep, and Mijailovich}]{Smith_2008}
Smith, D., Geeves, M., Sleep, J., Mijailovich, S., Oct. 2008. {Towards a
  Unified Theory of Muscle Contraction. 1: Foundations}. Ann. Biomed. Eng. 36,
  1624--1640.

\bibitem[{Steuerman et~al.(2004)Steuerman, Tseng, Peters, Flood, Jeppesen,
  Nielsen, Stoddart, and Heath}]{Steuerman:2004fx}
Steuerman, D.~W., Tseng, H.~R., Peters, A.~J., Flood, A.~H., Jeppesen, J.~O.,
  Nielsen, K.~A., Stoddart, J.~F., Heath, J.~R., Dec. 2004.
  {Molecular-Mechanical Switch-Based Solid-State Electrochromic Devices}.
  Angewandte Chemie 116~(47), 6648--6653.

\bibitem[{S{\"u}dhof(2013)}]{Sudhof:2013fw}
S{\"u}dhof, T.~C., 2013. Neurotransmitter release: The last millisecond in the
  life of a synaptic vesicle. Neuron 80~(3), 675--690.

\bibitem[{van Kampen(2001)}]{vanKampen:2001vs}
van Kampen, N., 2001. Stochastic Processes in Physics and Chemistry.
  North-Holland, Amsterdam.

\bibitem[{Wu et~al.(2016)Wu, Harne, and Wang}]{Wu:2015hi}
Wu, Z., Harne, R., Wang, K., Apr. 2016. {Exploring a modular adaptive
  metastructure concept inspired by muscles cross-bridge}. J. Intell. Mat.
  Syst. Struct. 27~(9), 1189--1202.

\bibitem[{Wu et~al.(2014)Wu, Harne, and Wang}]{Wu:2014dz}
Wu, Z., Harne, R.~L., Wang, K.~W., 2014. {Muscle-Like Characteristics With an
  Engineered Metastructure}. ASME 2014 Conference on Smart Materials, Adaptive
  Structures and Intelligent Systems, V002T06A019--V002T06A019.

\bibitem[{Xin and Lu(2016)}]{Xin:2016by}
Xin, F., Lu, T., Jun. 2016. {Tensional acoustomechanical soft metamaterials.}
  Sci Rep 6, 27432.

\bibitem[{Zwanzig(1988)}]{Zwanzig:1988vl}
Zwanzig, R., Apr. 1988. {Diffusion in a rough potential.} PNAS 85~(7),
  2029--2030.

\end{thebibliography}
\bibliographystyle{elsarticle-harv.bst}

\end{document}